\documentclass[aps,prb,twocolumn,groupedaddress,floatfix,showpacs,endfloats]{revtex4-1}

\usepackage{graphicx}
\usepackage{amsmath}
\usepackage{subcaption}
\usepackage{color}

\begin{document}


\title{Electrooptical Properties of Rydberg Excitons}


\author{Sylwia Zieli\'{n}ska-Raczy\'{n}ska}
\author{David Ziemkiewicz}
\email{david.ziemkiewicz@utp.edu.pl}
\author{Gerard Czajkowski}
\affiliation{Institute of Mathematics and Physics, UTP University
of Science and Technology, \\ Al. Prof. S. Kaliskiego 7, 85-789
Bydgoszcz, Poland.}


\date{\today}

\begin{abstract}
We show how to compute the electrooptical  functions (absorption,
reflection, and transmission) when Rydberg Exciton-Polaritons
appear, including the effect of the coherence between the
electron-hole pair and the electromagnetic  field. With the use of
Real Density Matrix Approach  numerical calculations applied for
Cu$_2$0 crystal are performed.  We also examine in detail and
explain the dependence of the resonance displacement on the state
number and applied electric field strength. We report a good
agreement with recently published experimental data.
\end{abstract}


\maketitle
\section{Introduction}
 The new avenue in modern semiconductor physics has been opened by outstanding experiment performed recently by Kazmierczuk \emph{et al}
 \cite{Kazimierczuk} who detected the large quasi particles known as Rydberg excitons in natural crystal of  copper oxide.
 They have observed absorption lines associated with excitons of principal quantum numbers as large as $n=25$.
  One could expect that Rydberg excitons would have been described, in analogy to Rydberg atoms, by Rydberg
   series of  hydrogen atom, but it has turned out that this generic method of description should have been revised.
   This is due to the fact that a size of a huge quasi particle, which in fact is a Rydberg exciton with high \emph{n}, has a
   diameter more then two micrometers, which is much larger then  the wavelength of light needed to create this exciton.
   Several theoretical approaches to calculate optical properties of Rydberg excitons have been presented \cite{Hofling}$^-$\cite{assmann_2016}.
   Schweiner \emph{et al} \cite{Schweiner} developed calculations of of the absorption spectrum on the ground of Toyozawa theory and
    calculated the main parameters for excitonic absorption line for yellow exciton series and emphasized that central-cell corrections have a
     major influence on the linewidth of the $2P$-exciton state.  In our recent paper \cite{Zielinska.PRB} we have proposed the method based
     on the Real Density Matrix Approach (RDMA) to obtain the analytical expressions for the optical functions of
     semiconductor crystals, including a high number of Rydberg excitons, taking into account the effect of
     anisotropic dispersion and the coherence of the electron and hole with the radiation field.

  It is expected that the natural direction of development interest in Rydberg excitons is focused on
  Stark effect in such  systems because this phenomenon
   may be used for the optical manipulations of excitons if there is an efficient coupling between the radiation
  field and excitonic systems far from the band edge. The copper oxide is a perfect candidate for such observations because due to high binding energy of orders of hundred meV and due to their large size, Rydberg excitons in Cu$_2$0 can exhibit very large electric dipole moments.
These features provide that this system is appropriate to observe
Stark effect experimentally. In semiconductors where the Wannier
excitons have a small binding energy (as, for example, GaAs), the
main effect of the applied electric field is the Stark shift of
the excitonic resonances and changes in their oscillator strengths
(see, for example, \cite{EPJ}, \cite{RivistaGC} for a review). In
Cu$_2$O which is now the main semiconductor where Rydberg excitons
are observed, even relatively high excitonic states have a binding
energy which is larger than the corresponding ionization energy.
Thus the excitonic character of the spectra is conserved, but new
phenomena, as for example the appearance of symmetry forbidden
states, with positions dependent on the applied field strength
and on the state number, are observed~\cite{Schoene,Verhandl}.

Actually,  one of the aims of our theoretical paper is to
extend the method presented in ref.~\cite{Zielinska.PRB}, which
allowed to describe optical properties of Rydberg excitons, in
order to obtain electrooptic functions (susceptibility,
 absorption reflection and transmission). Our approach has general character because it works for any exciton angular momentum number and for an arbitrary electric field.
    In particular, we derive
 an analytical expression for the electrosusceptibility, from
 which other electrooptical functions can be obtained.  Since the
 electric field effects, increasing with the applied field strength
 and the state number, compete with the decreasing oscillator
 strength, we are able, having analytical expressions, to indicate
 the optimal excitation interval to observe the electrooptical
 effects. We also indicate the impact of the finite crystal size
 on the shape of the spectra, which was overlooked in the previous
 considerations.
  Therefore our predictions should be of interest for
experimentalists.

  The motivation of our considerations is also connected with with potential application of
   Rydberg excitons
   as solid-state switches. Due to their unusual features: long lifetimes, strong dipolar
   interactions and huge-size, they are expected to be implemented in quantum information technology.
 Kazimierczuk \emph{et al}~\cite{Kazimierczuk} observed  Rydberg blockade RB, which consists
 in reduction of excitonic absorption accompanied by increasing laser power
 for lines associated with large n, what means that only limited amount of
 Rydberg excitons is permitted in a well-localized space of the crystal.
 The idea of using dipolar Rydberg interaction to implement RB bases on
 the fact that in an ensemble of particles coupled by long-range dipolar interactions, only one particle can be excited at  given time.  The blockade originate  from dipole-dipole interactions between Rydberg excitons unnecessary with the same \emph{n} and is strongly influenced by their separation.
 This effect offers exciting possibilities for manipulating quantum bits
 stores in a single collective excitation in mezoscopic ensembles or for
 realizing scalable quantum logic gates and one implemented in solids
 would bring a lot of advantages for quantum information, for constructing
  all-optical switches and single-photon logic devices. Moreover, it is
  essential to have  an additional mechanism for switching the dipole-dipole
  interaction, which in fact can be tuned on and off by Stark effect, therefore it is worth to
   go into details of the Stark effect in Rydberg excitons.

Our paper is organized as follows. In  Sec. \ref{density.matrix}
 we present the assumptions of considered model and solve the
 constitutive equation which give an analytical expression for the electrosusceptibility.
 We also use the obtained expression to compute
 the effective dielectric function, thanks to which  the
  electrooptical functions (reflectivity, transmissivity, and
  absorption) are derived (section \ref{eofunctions}). Next, in Sec.~\ref{results}, the
electrooptic functions  are numerically analyzed for
 Cu$_2$O crystal for the purpose of realistic implementation of presented method. We examine in details changes of
 both real and imaginary part of electrosusceptibility, reflectivity and transmission under the influence of electric field.
   In Sec. \ref{conclusions} we draw conclusions of the model studied in this paper and  we indicate the optimal range of energy for which Rydberg excitons could be experimentally observed.

\section{Density matrix formulation}\label{density.matrix}
Wannier-Mott excitons, treated as hydrogen-like particles, due to
their small binding energy, are very receptive to the action of
external fields (electric and/or magnetic). The external fields
remove the degeneration of the excitonic energy levels and enhance
the optical effects. Such effects were observed in the case of
Rydberg excitons in Cu$_2$O~\cite{Kazimierczuk},\cite{Thewes}. In
what follows we describe the electrooptic properties of systems
where the Rydberg excitons appear. As was recently shown in
ref.~\cite{Zielinska.PRB}, the so-called real density matrix
approach is very effective in describing the optical properties of
Rydberg excitons. This approach was used in the past for the
description of electrooptical effects (see, e.g., ref. \cite{EPJ} and the references therein). We show below, that the specific
properties of Rydberg excitons require a reformulation of the
methods used in the past. As in ref.~\cite{Zielinska.PRB}, we do
not enter into the quantum-mechanical explanation of the valence
band structure of the Cu$_2$O. This explanation is given in
details in the recent paper by Schweiner \emph{et
al}~\cite{Schweiner_2016}. Here we treat the band structure and
the related parameters as known, and
 use the scheme of ref.~\cite{Zielinska.PRB} for the
situation, when the constant external electric field $\textbf{F}$
is applied in the $z$ direction. The presented method starts with
the constitutive equations, which have the form (for example,
\cite{RivistaGC},\cite{StB87})
 \begin{eqnarray}\label{constitutiveeqn}
 \dot{Y}(\textbf{R},\textbf{r})&=&(-{\rm
 i}/\hbar)H_{eh}{Y}(\textbf{R},\textbf{r})+e\textbf{F}\textbf{r}-{\mit\Gamma}{Y}(\textbf{R},\textbf{r})\nonumber\\&+&({\rm i}/\hbar)\textbf{E}(\textbf{R})\textbf{M}(\textbf{r}),
 \end{eqnarray}
where $Y$ is the bilocal coherent electron-hole amplitude (pair
wave function), ${\bf R}$ jest is the excitonic center-of-mass
coordinate, $\textbf{r}=\textbf{r}_e-\textbf{r}_h$ the relative
coordinate, $\textbf{M}(\textbf{r})$ the smeared-out transition
dipole density, ${\bf E}({\bf R})$ is the electric field vector of
the wave propagating in the crystal. The coefficient $\mit\Gamma$
in the constitutive equation  represents dissipative processes. We can expect a significant temperature-dependence of the
spectra; microscopic analysis of damping parameters, which are the main temperature-dependent factors, requires future studies and will not be considered explicitly in this paper. The interaction with phonons and their role in determining the line shape, discussed recently by Schweiner at al (ref. \cite{Schweiner}) who have considered possible causes of line broadening, goes into the field of nonlinear optics and was in the past considered in the framework of the RDMA, for example by Schl\"osser (ref. \cite{Schlosser}) or in the case of EIT, in ref.  \cite{EPJGC}. In this paper we do not consider the interaction with phonons, and take the damping coefficients as phenomenological constants.

 The smeared-out transition dipole density ${\bf M}({\bf
r})$ is related to the bilocality of the amplitude $Y$ and
describes the quantum coherence between the macroscopic
electromagnetic field and the interband transitions. The two-band
Hamiltonian $H_{eh}$ includes the electron- and hole kinetic
energy terms, the electron-hole interaction potential and the
confinement potentials. For details about the Hamiltonian see, for
example, \cite{Zielinska.PRB}. The coherent amplitude $Y$ defines
the excitonic counterpart of the polarization
\begin{equation}\label{polarization}
\textbf{P}(\textbf{R})=2 \int {\rm d}^3 r~
\hbox{Re}~\left[\textbf{M}(\textbf{r})Y(\textbf{R},\textbf{r})\right],
\end{equation}
which is than used in the Maxwell field equation
\begin{equation}\label{Maxwell}
c^2\nabla_R^2
\textbf{E}-\underline{\underline{\epsilon}}_b\ddot{\textbf{E}}(\textbf{R})=\frac{1}{\epsilon_0}\ddot{\textbf{P}}(\textbf{R}),
\end{equation}
with the use  of the bulk dielectric tensor
$\underline{\underline{\epsilon}}_b$ and the vacuum dielectric
constant $\epsilon_0$. In the present paper we solve the equations
(\ref{constitutiveeqn})-(\ref{Maxwell}) with the aim to compute
the electrooptical functions (reflectivity, transmission, and
absorption) for the case of Cu$_2$O. In the following we will start
with considering the bulk situation, where the center-of-mass motion is
decoupled from the relative electron-hole motion and given by the
term $\exp({\rm i}\textbf{k}\,{\textbf{R}})$ with the wave vector
$\textbf{k}$ resulting, in general, from the polariton dispersion
relation~\cite{Zielinska.PRB}. We also assume the harmonic time
dependence $\propto \exp(-{\rm i}\omega t)$. This assumptions
allow to
 calculate the dielectric susceptibility. This
will be achieved in
 by expanding the coherent amplitudes $Y$ in
terms of eigenfunctions of the Hamiltonian $H_{eh}$.  Let us note
that the solution of the Schr\"{o}dinger equation
\begin{equation}\label{schroed}
H_{eh}\Psi+V(\textbf{r})\Psi=E\Psi,\quad
V(\textbf{r})=e\textbf{Fr},
\end{equation}
can be obtained only in an approximative way (perturbation
calculus, variational method, matrix diagonalization etc.).
Considering the cases of Cu$_2$O, when the applied field is of the
order of 10 V/cm (\cite{Thewes}), we can compare the magnitude of
the electron-hole pair attractive energy ($E_n=-R^*/n^2$ in the
isotropic effective masses approximation) and the electric field
energy $E_{field}=eFa^*_n, a^*_n=n^2a^*$. For $n=16$ one has
$E_{Coulomb}=0.39~\hbox{meV}$ and $E_{field}=0.38~\hbox{meV},
\hbox{when}~F=15~\hbox{V/cm}$ (\cite{Thewes}). Thus the excitonic
character of the spectra prevails and the applied electric field
can be considered as perturbation. It
is clear that, when external fields are applied, the full
diagonalization of field- and band-mixing effects is more
adequate to describe the optical properties, in particular
when polarization dependence is considered. However, in
the RDMA the band parameters, as e.g. the effective
masses, are considered as field independent, and
the fields are treated as perturbation operators. Such approach
was merely applied in the past (for a recent review see \cite{EPJ}), for various nanostructures and field
orientations, and was justified by the agreement with experimental
data. The considered approximation is also
justified by the fact, that the applied field strengths are
much below the critical values for the fields (the ionization
field for the electric field and the critical magnetic
field). In the case of Cu$_2$0 the ionization field is of the order
of 10$^6$ V/cm, compared to the applied 15 or even 50 V/cm.

We assume the solution of the
Eq. (\ref{schroed}) in form of the solutions of an anisotropic
Schr\"{o}dinger equation $\varphi_{n\ell m}$ (see
Appendix~\ref{Appendix A} for details)
\begin{equation}\label{eigenf}
\varphi_{n\ell m} (\textbf{r})=R_{n\ell m}(r)Y_{\ell
m}(\theta,\phi),
\end{equation}
where\begin{eqnarray}\label{radialfinala} R_{n\ell
m}(r)&=&\left(\frac{2\eta_{\ell
m}}{na^*}\right)^{3/2}\sqrt{\frac{(n-\ell-1)!}{2n(n+\ell)!}}\left(\frac{2\eta_{\ell
m}r}{na^*}\right)^\ell\nonumber\\
&&\times L_{n-\ell-1}^{2\ell+1}\left(\frac{2\eta_{\ell
m}r}{na^*}\right) e^{-\eta_{\ell m}r/na^*},
\end{eqnarray}
with $\eta_{\ell m}$ defined by (\ref{etaellm}), and the Laguerre
polynomials $L_n^\alpha(x)$ (for example, \cite{Grad})
\begin{eqnarray}\label{laguerre}
L^\alpha_n(x)&=&\frac{1}{n!}e^xx^{-\alpha}\frac{{\rm d}^n}{{\rm
d}x^n}\left(e^{-x}x^{n+\alpha}\right)\nonumber\\&=&\sum\limits_{m=0}^n(-1)^m{n+\alpha\choose
n-m}\frac{x^m}{m!},
\end{eqnarray}
$Y_{\ell m}$ being the spherical harmonics. The energy eigenvalues
relate to the eigenfunctions (\ref{eigenf}) have the form (see Eq.
(\ref{energies}))
\begin{equation}\label{energies1}
E_{n\ell m}=-\frac{\eta_{\ell m}^2}{n^2}R^*,
\end{equation}
where $n=1,2,\ldots, \ell=0,1,2,\ldots n-1, m=-\ell,-\ell+1,\ldots
+\ell$. We see that the mass anisotropy removes the degeneracy
with respect to the quantum number $\ell$, so that in this
approach the higher order excitons $P$, $D$, $F$ etc. appear.
 When the
electric field is directed along the \emph{z}-axis, the
perturbation operator $V$ has the form
\begin{equation}\label{Vzet}
V=eFz=eFr\cos\theta.
\end{equation}
We look for the solutions of Eq. (\ref{constitutiveeqn}) in the
form
\begin{equation}\label{expansion}
Y=\sum\limits_{n\ell m}c_{n\ell m}R_{n\ell m}(r)Y_{\ell
m}(\theta,\phi).
\end{equation}
Inserting the above expansion into (\ref{constitutiveeqn}) we
obtain the following system of equations for the expansion
coefficients $c_{n_1\ell_1m_1}$ (for details, see Appendix B)
\begin{eqnarray}\label{basic}
X_{n_1 \ell_1
m_1}&=&c_{n_1\ell_1m_1}W_{n_1\ell_1m_1}+\sum\limits_nc_{n\ell_1-1
m_1}V^{(n)}_{\ell_1-1\ell_1m_1}
\nonumber\\&+&\sum\limits_nc_{n\ell_1+1
m_1}V^{(n)}_{\ell_1\ell_1+1m_1}
\end{eqnarray}
where
\begin{eqnarray}\label{V12}
\label{V1}&&V^{(nn_1)}_{\ell_1-1\ell_1m_1}\nonumber=eF\sqrt{\frac{\ell_1^2-m_1^2}{4\ell_1^2-1}}\int\,r^2{\rm
d}r\,R_{n_1\ell_1m}rR_{n\ell_1-1m},\nonumber\\
\label{V2}&&V^{(nn_1)}_{\ell_1\ell_1+1m_1}=\nonumber\\&&=eF\sqrt{\frac{(\ell_1+1)^2-
m_1^2} {(2\ell_1+1)(2\ell_1+3)}}\int\,r^2{\rm
d}r\,R_{n_1\ell_1m}rR_{n\ell_1+1m},\quad\nonumber\\
 &&X_{n_1 \ell_1 m_1}=\mathcal E\int {\rm
d}\Omega\int r^2{\rm d}r
Y_{\ell_1
m_1}R_{n_1m\ell_1}M(r,\theta,\phi),\nonumber\\
 &&W_{n\ell m}=E_g+E_{n\ell
m}+\frac{\hbar^2}{2M_z}k^2-\hbar\omega-{\rm
i}{\mit\Gamma}=\nonumber\\&&=E_{Tn\ell m}-E+\frac{\hbar^2}{2M_z}k^2-{\rm
i}{\mit\Gamma},
\end{eqnarray}
where $\mathcal E$ denotes the amplitude of the electric field.
 In all calculations we will use only the above matrix elements with $n=n_1$, denoting them by
  $V^{(n)}_{\ell_1-1\ell_1m_1}, V^{(n)}_{\ell_1\ell_1+1m_1}.$ This
  is an approximation, which can be justified as follows. The spacing between the Rydberg states,
   at least for the states $ n=2,..,7$ considered in this paper, is of the order of a few meV. Taking, for simplicity, Rydberg equal to 100 meV,
one has the spacings (taking $\ell=0,m=0$ states) (meV)
$E_3-E_2=14, E_4-E_3=4.75, E_5-E_4 =2.25$, etc. On the other hand,
the matrix elements \emph{V}, collected in Table
\ref{matrix.elements} and being the measure of the splitting
between the Stark levels with the same principal number, are of
the order between $10^{-3}$ and $10^{-1}$ meV, so that there are
much smaller than the distances between the exciton states.
Obviously, one should notice that the distances between the
excitonic states decrease with the increasing number n, whereas
the Stark splittings increase, and at a certain number the Stark
splittings are greater than the spacing between the Rydberg
states. The indication is that for higher numbers \emph{n} one
should take into account the interaction (in other words the
matrix elements \emph{V}) between different states, and not only
within the same state. Besides, the method applied is not exactly
the perturbation calculus, rather the matrix diagonalization, so
its validity is not restricted by the value of the applied field.

\begin{figure}
\includegraphics[width=0.65\linewidth]{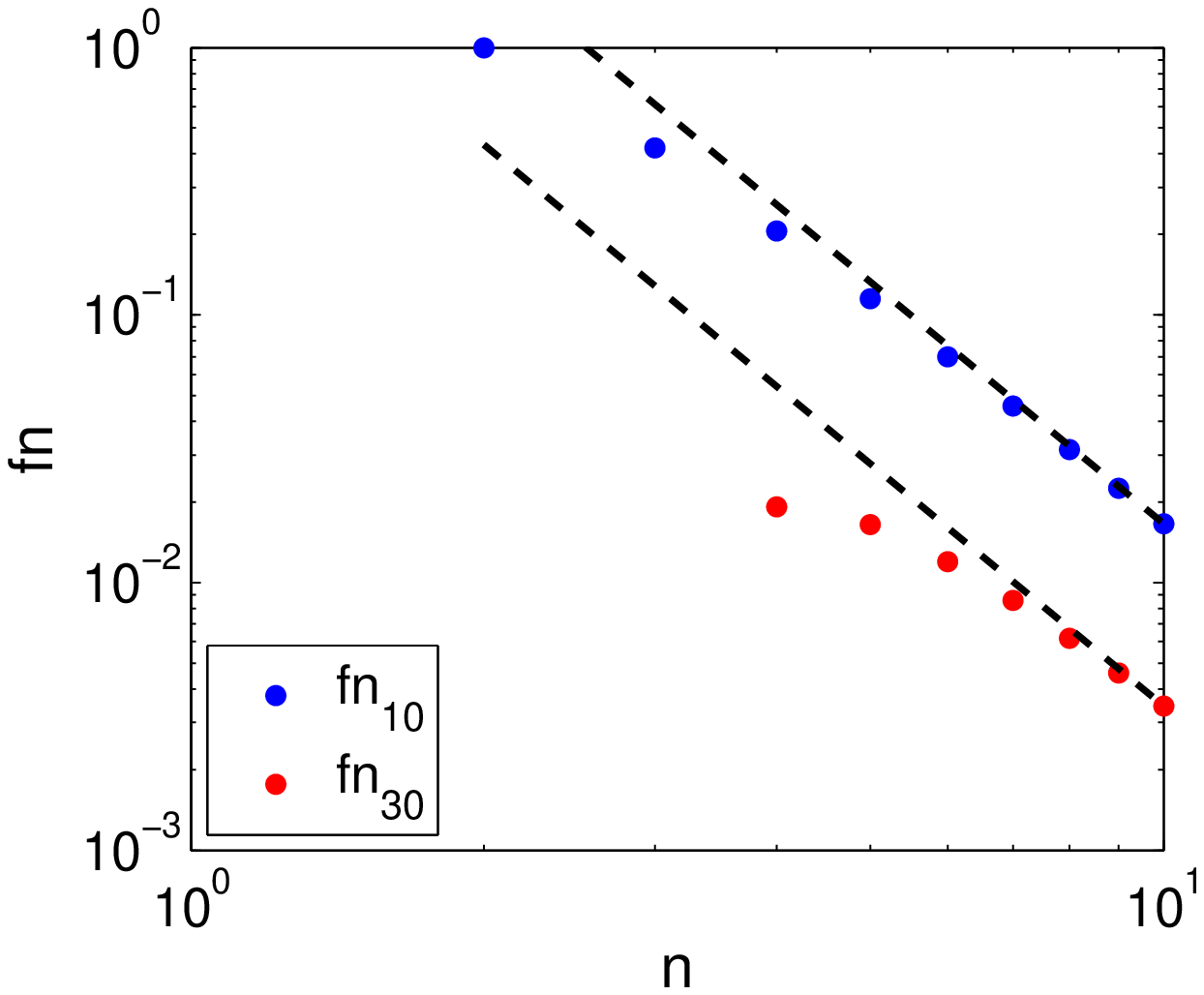}
\caption{Oscillator strengths as a functon of exciton number $n$. Logarithmic scale is applied. Dashed line marks the linear regression for $n^{-3}$ relation.}\label{Figfnn}
\end{figure}

  We put the coherent
amplitudes (\ref{expansion}) into the equation
(\ref{polarization}), from which, when the center-of-mass motion
is decoupled, one can obtain the susceptibility from the relation
$\textbf{P}=\epsilon_0\chi(\omega, \textbf{k})\textbf{E}$. The
dipole density vectors $\textbf{M}$ should be chosen appropriate
for \emph{P}- or \emph{F}- excitons, and we obtain (see also
\cite{Zielinska.PRB})
\begin{eqnarray}\label{susceptibility}
\chi(\omega,\textbf{k})&=&\Delta_{LT}^{(2)}\sum\limits_{n=2}^NC_{n10}f_{n1}
+\Delta_{LT}^{(2)}\sum\limits_{n=4}^Nf_{n3}C_{n30}
\end{eqnarray}
where for \emph{ P} excitons
\begin{eqnarray}\label{cc210}
C_{210}&=&\frac{W_{200}}{W_{200}W_{210}-\left(V_{010}^{(2)}\right)^2},\nonumber\\
\label{c310}C_{310}&=&\frac{W_{300}W_{320}}{W_{300}\left[W_{310}W_{320}-\left(V_{120}^{(3)}\right)^2
\right]-\left(V_{010}^{(3)}\right)^2W_{320}}.\qquad
\end{eqnarray}
For $n\geq 4$ we take the same expression as (\ref{cc210})
\begin{eqnarray}
C_{n10}&=&\frac{W_{n00}}{W_{n00}W_{n10}-\left(V_{010}^{(n)}\right)^2}.\nonumber
\end{eqnarray}
For \emph{F} excitons when $\hbox{for}~n_1\geq 4$, taking into
account $\ell=0,1,2,3$, one has (see also (\ref{cn130}))
\begin{eqnarray}\label{cn130a}
C_{n_130}&=&\frac{W_{n_120}W_{n_110}W_{n_100}}{\Delta}\nonumber\\&-&\frac{W_{n_120}\left(V_{010}^{(n_1)}\right)^2
-W_{n_100}\left(V_{120}^{(n_1)}\right)^2}{\Delta},\nonumber\\
\Delta&=&W_{n_130}W_{n_120}W_{n_110}W_{n_100}\nonumber\\&&-W_{n_130}\left[W_{n_120}\left(V_{010}^{(n_1)}\right)^2
-W_{n_100}\left(V_{120}^{(n_1)}\right)^2\right]\nonumber\\
&&-\left(V^{(n_1)}_{230}\right)^2\left[W_{n_110}W_{n_100}-\left(V_{010}^{(n_1)}\right)^2\right].
\end{eqnarray}
For $F$ excitons, when $n\geq 5$, we can extend the basis taking
$\ell=4,3,2,1,0$ and $m=0$, obtaining the expressions
(\ref{cn130ext}). In the above formulas $\Delta_{LT}^{(2)}$
denotes the longitudinal-transverse splitting energy. The explicit
form of the oscillator strengths $f_{n1}, f_{n3}$ for the
isotropic case can be found in ref.~\cite{Zielinska.PRB} The oscillator strength $f_{n1}$ associated with P excitons is one order of magnitude greater than $f_{n3}$. Both values are roughly proportional to $n^{-3}$, especially for higher values of n (see Fig. \ref{Figfnn}).
The quantities of the type $W_{n00}$,$W_{n20}$ which enter into the above equations, correspond to S and D excitons. The matrix elements $V^{(n)}_{\ell\ell'm}$ are calculated in Appendix~\ref{Appendix C}.

\section{Reflection and transmission spectra}\label{eofunctions} In the previous
considerations we treated the semiconductor crystal as unbounded.
The real situation is different due to  crystal  finite size in
all directions.
Practically, the confined size in only one arbitrary chosen direction is considered and
 usually this direction is the same as electromagnetic wave vector.
Concerning the experiments with Cu$_2$O one should notice that the
dimension of the crystals examined experimentally exceeds the
electromagnetic wave length, therefore the use of the
long-wave-approximation is not well justified so we will compute
the optical functions such as the transmissivity and reflectivity
taking into account the finite crystal size and finite wavelength.
We will obtain analytic expressions for the optical functions.
These expressions will also include the impact of the applied
constant electric field. In the description of the optical
properties of excitons in finite semiconductors the excitonic Bohr
radius plays an important role. Near the semiconductor surfaces
there are layers where the excitons are created (or destroyed),
the so-called exciton-free layers ("dead layers"). Mostly it is
assumed that their thickness amounts to 2-3 excitonic Bohr radii.
In the case of GaAs it gives about 30 nm. The excitons and related
to them polaritons are formed in the remaining volume ("bulk") of
the crystal, and are responsible for the bulk susceptibility. The
junction of layers with different dielectric properties is a
complicated task and many works on this topic, including the
so-called ABC problem, have been done over the past decades (for
review see, for example,~\cite{RivistaGC,Bassani_2003,Agran2009}). When we consider a particular case of GaAs
thin layer of the thickness 150 nm, and the relevant excitonic
Bohr radius is about 15 nm, then two excitonic
Bohr radii correspond to 20 \% of the crystal size.
When we consider a Cu$_2$O slab, the situation is quite different.
For a Cu$_2$O crystal of the size 30 $\mu$m \cite{Kazimierczuk},
\cite{Thewes} even the exciton state with \emph{n}=25 has the
extension of about 0.6 $\mu$m, so that the two Bohr radii
correspond to 4 \% of the crystal size. This means that, in the
first approximation, we can neglect the dead layer effects.  It
does not mean that the dead layer and polariton effects are not
important, as they can shift the resonance positions and affect the oscillator strengths (see also the discussion in ref. \cite{Schweiner}). The
problem is that, when taking into account 25 excitonic states, we
have at least 50 polaritonic waves (including the in- and outgoing
polariton waves) so that the methods applied for the III-V and
II-VI compounds (for example, ref. \cite{CBT96})  cannot be
applied for the case under consideration. This aspect requires
future studies and will not be explicitly considered in this
paper. Moreover it should be mentioned that different aspect of semiconductors'
geometry was considering by Schweiner \emph{et al} \cite{Schweiner_2016} who have developed the method of investigation excitonic spectra taking into account the discrepancy of valence and conduction bands from parabolic shapes as well as their degeneracy and possible anisotropy.

The formation of excitons can be considered as a fast
process leading to an effective dielectric function
\begin{equation}\label{epsiloneffective}
\epsilon_{eff}=\epsilon_b+\chi=\epsilon_b+\chi_1+{\rm i}\,\chi_2,
\end{equation}
with the excitonic susceptibility defined in Eq.
(\ref{susceptibility}). Thus the electromagnetic wave in the
crystal propagates in a medium characterized by the effective
dielectric function. The crystal under consideration will be
modeled by a slab with infinite extension in the $xy$-plane and
the boundary planes $z=0, z=L$. With the sake of simplicity, the
slab is located in vacuum. An monochromatic, linearly polarized
electromagnetic wave propagates along $z$ axis. Its
electric field is given by
\begin{equation}
{\bf E}=(E_x, 0, 0),\qquad E_x=E_{in}e^{{\rm i}k_0z-{\rm i}\omega
t}\,,
\end{equation}
\noindent where for vacuum
\begin{equation}
k_0=\frac{\omega}{c},
\end{equation}
\noindent $\omega$ being the frequency, and $c$ the velocity of
light. It is well known, the energy of the propagating wave will
be divided into reflected and transmitted wave. The reflectivity,
transmissivity, and absorption will be obtained from the
relations

\begin{eqnarray}\label{wzoryfcjeoptyczne}
R=\left|\frac{E(0)}{E_{in}}-1\right|^2,&\qquad&
T=\left|\frac{E(z=L)}{E_{in}}\right|^2,\\
A&=&1-R-T,\nonumber
\end{eqnarray}
\noindent 
where $E(z)$ is the $x$-component of the wave electric field inside the
crystal.

In the simplest approximation, neglecting the carrier confinement
effects leading to the above mentioned ABC problem, we can use the
effective dielectric function (\ref{epsiloneffective}) and the
resulting effective refractive index
\begin{eqnarray}
n&=&\sqrt{\epsilon_{eff}}=n_1+{\rm i}\,n_2,\nonumber\\
n_1&=&\hbox{Re}\;n\approx\sqrt{\epsilon_b+\chi_1},\\
n_2&=&\hbox{Im}\;n\approx \frac{\chi_2}{2n_1}.\nonumber
\end{eqnarray}
Then the reflectivity results from the standard formula
\begin{equation}\label{Rcoeff}
R=\left|\frac{1-n}{1+n}\right|^2=\frac{\left(1-n_1\right)^2+n_2^2}{\left(1+n_1\right)^2+n_2^2}.
\end{equation}

Regarding  the exceptional  experiments by Kazimierczuk \emph{et
al}~\cite{Kazimierczuk} and Thewes \emph{et al}.~\cite{Thewes} we can use
the model of the multiple reflection and  in the lowest order we get the following expression
describing transmission
\begin{eqnarray}\label{transmission}
&&T=\frac{16\vert n\vert^2}{\vert(1+n)^2\vert^2}\;e^{-\alpha
L}=\nonumber\\&&=\frac{16\left(n_1^2+n_2^2\right)}{\left[\left(1+n_1\right)^2-n_2^2\right]^2+4n_2^2\left(1+n_1\right)^2}\;e^{-\alpha
L}.
\end{eqnarray}
Here
\begin{equation}\label{alpha}
\alpha=2\frac{\hbar\omega}{\hbar c}\hbox{Im}\,n
\end{equation}
denotes the absorption coefficient.
\section{Results of specific calculations}\label{results}
\begin{figure}[htb!]
\includegraphics[width=1\linewidth]{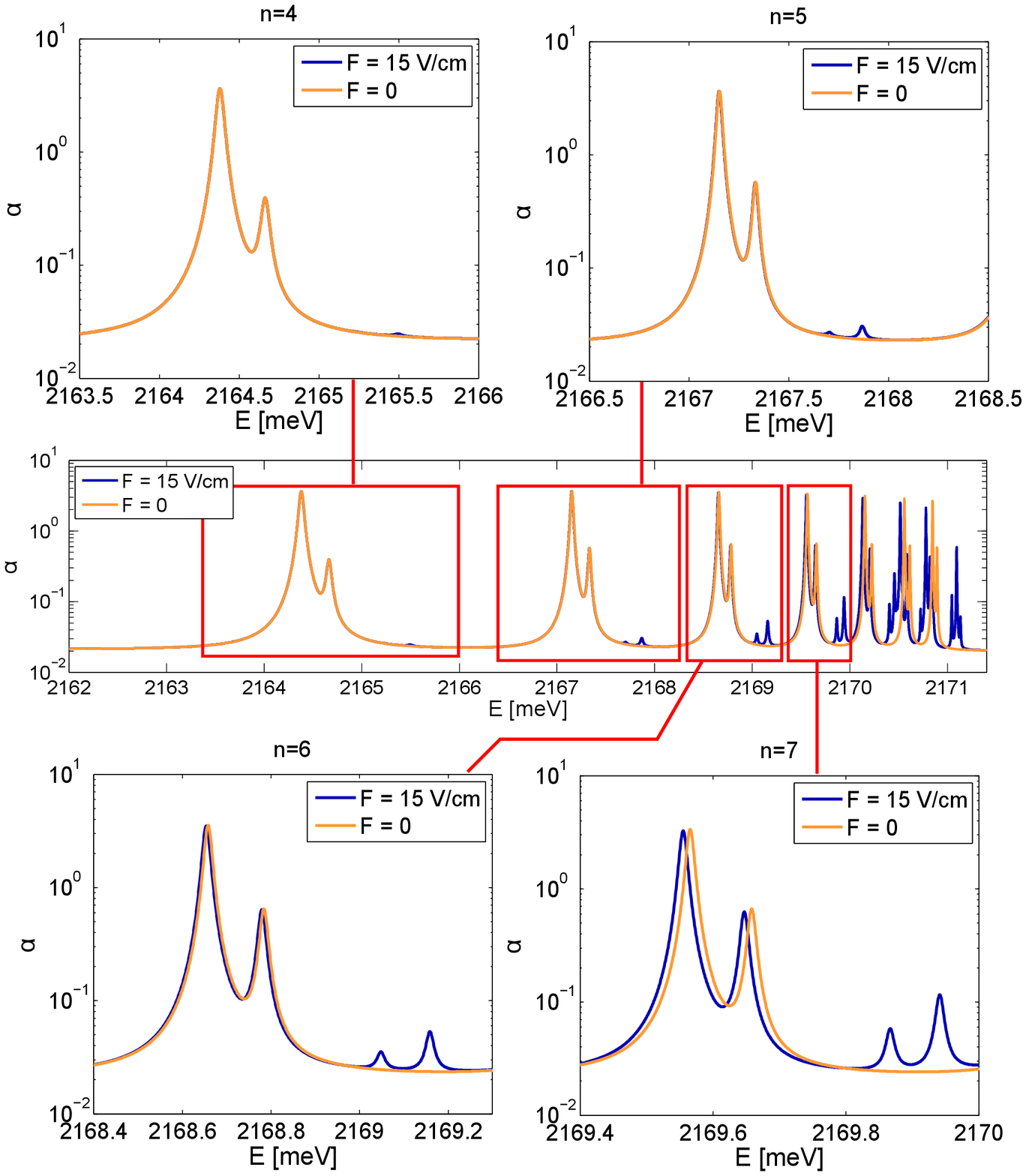}
\caption{The bulk electroabsorption of a Cu$_2$O crystal
calculated from the imaginary part of the susceptibility in the
energetic region of $n=3-10$ excitonic states, for the electric
field strengths $F=15~\hbox{V/cm}$ and $F=0$. The logarithmic
scale is applied. Insets show the absorption spectrum near
selected states.}\label{fig1_1}
\end{figure}

We have performed numerical calculations of electrooptical
functions (absorption, reflectivity, and transmissivity) for the
Cu$_2$O crystal having in mind the experiments by Thewes \emph{et
al}~\cite{Thewes}, and Sch\"{o}ne \emph{et al}~\cite{Schoene}.
First, using the obtained expression for the susceptibility
(\ref{susceptibility}-\ref{cn130a}), we have calculated the
electroabsorption, taking into account the lowest $n=2-10$
excitonic states. The parameters we used are the energies
$E_{n\ell m}$, the gap energy $E_g$, the L-T energy
$\Delta_{LT}^{(2)}$, and the dissipation parameter $\mit\Gamma$.

\begin{figure}[h]
\centering
\includegraphics[width=1\linewidth]{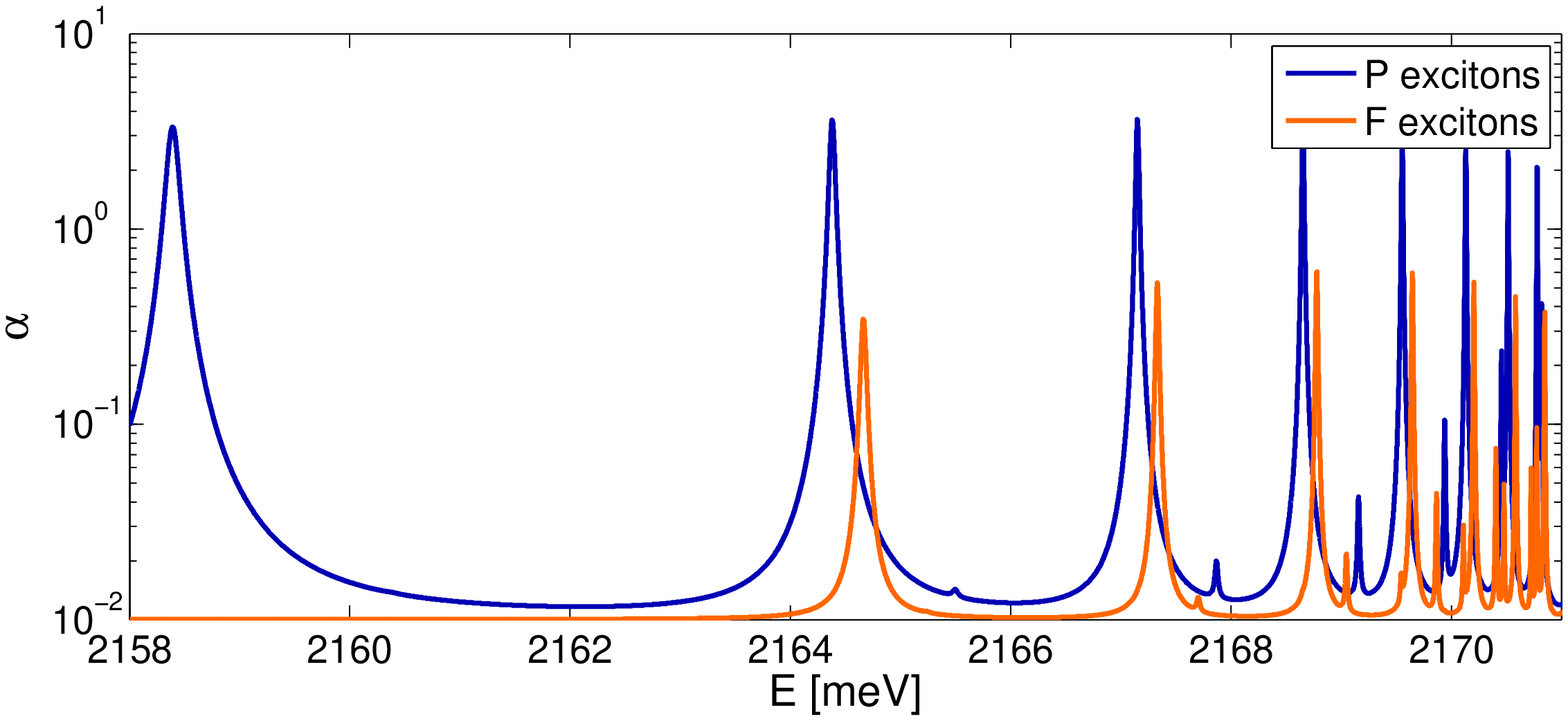}
\caption{The  bulk electroabsorption of a Cu$_2$O crystal,
in the energetic region of $n=3-10$ excitonic states, for the
electric field $F=15~\hbox{V/cm}$.  The logarithmic scale is
applied.  The contributions of \emph{P} and \emph{F} excitons
shown separately.} \label{Fig3ab}
\end{figure}

\begin{figure}[h]
\centering
\includegraphics[width=1\linewidth]{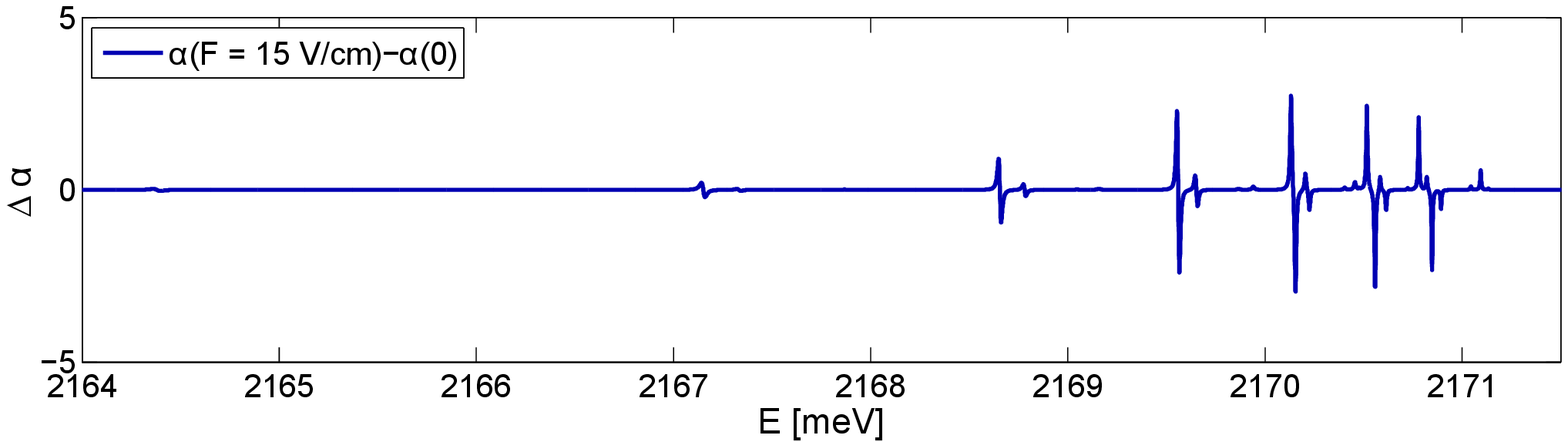}
 \caption{ \small The difference
$\Delta\alpha=\alpha(F)-\alpha(0)$, for $F=15~\hbox{V/cm}$ and in
the range of $n=4-10$ excitonic states} \label{Figdiff4}
\end{figure}

\begin{figure}[h]
\includegraphics[width=1\linewidth]{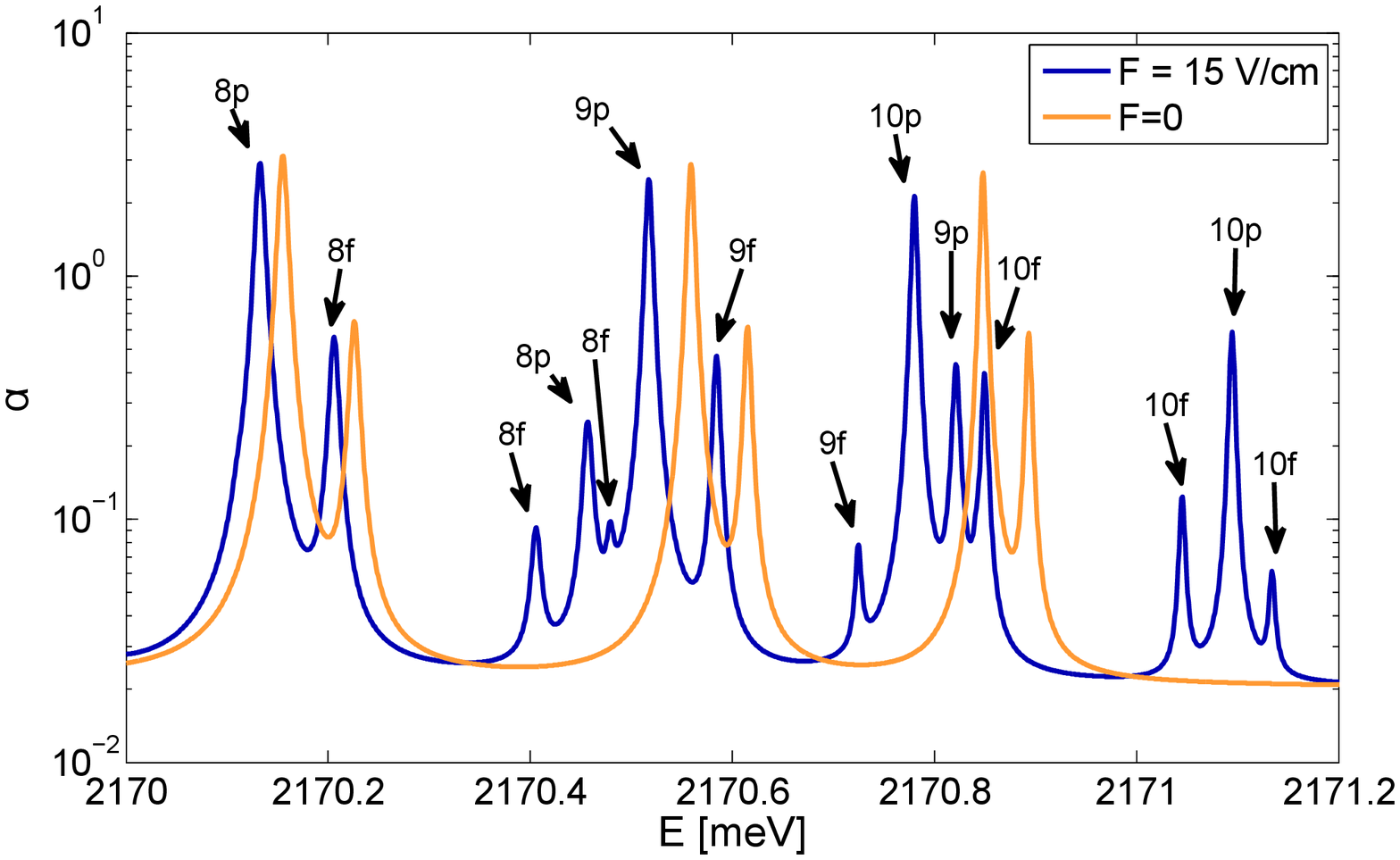}
\hfill
 \caption{ \small The bulk electroabsorption of $Cu_2O$ crystal calculated from imaginary part of susceptibility in
the energetic region of $n=8-10$ excitonic states, for two
electric field strengths $F=15~\hbox{V/cm}$ and $F=0$. The
logarithmic scale is applied. There is some overlap in the
identified states.} \label{Figasess}
\end{figure}

\begin{figure}[h]
\includegraphics[width=1\linewidth]{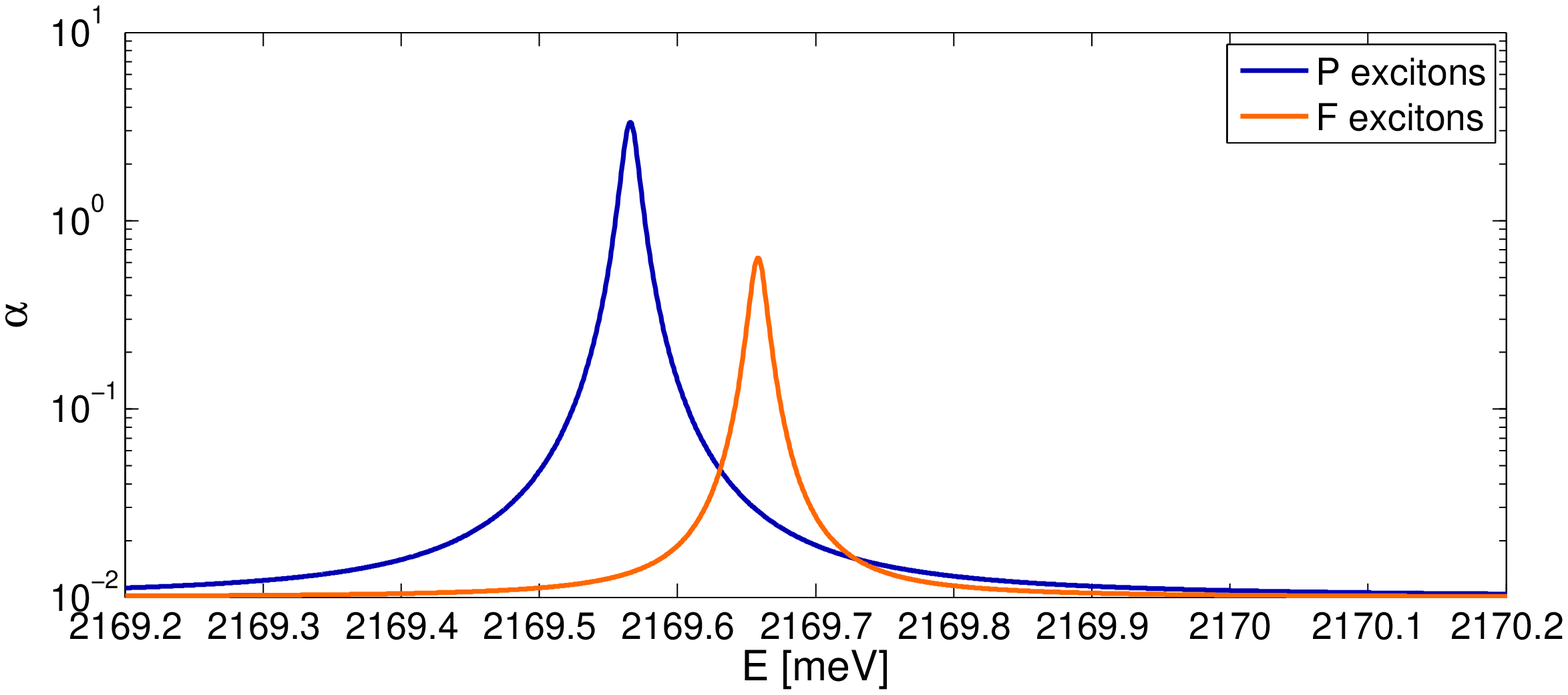}a)
\hfill
\includegraphics[width=1\linewidth]{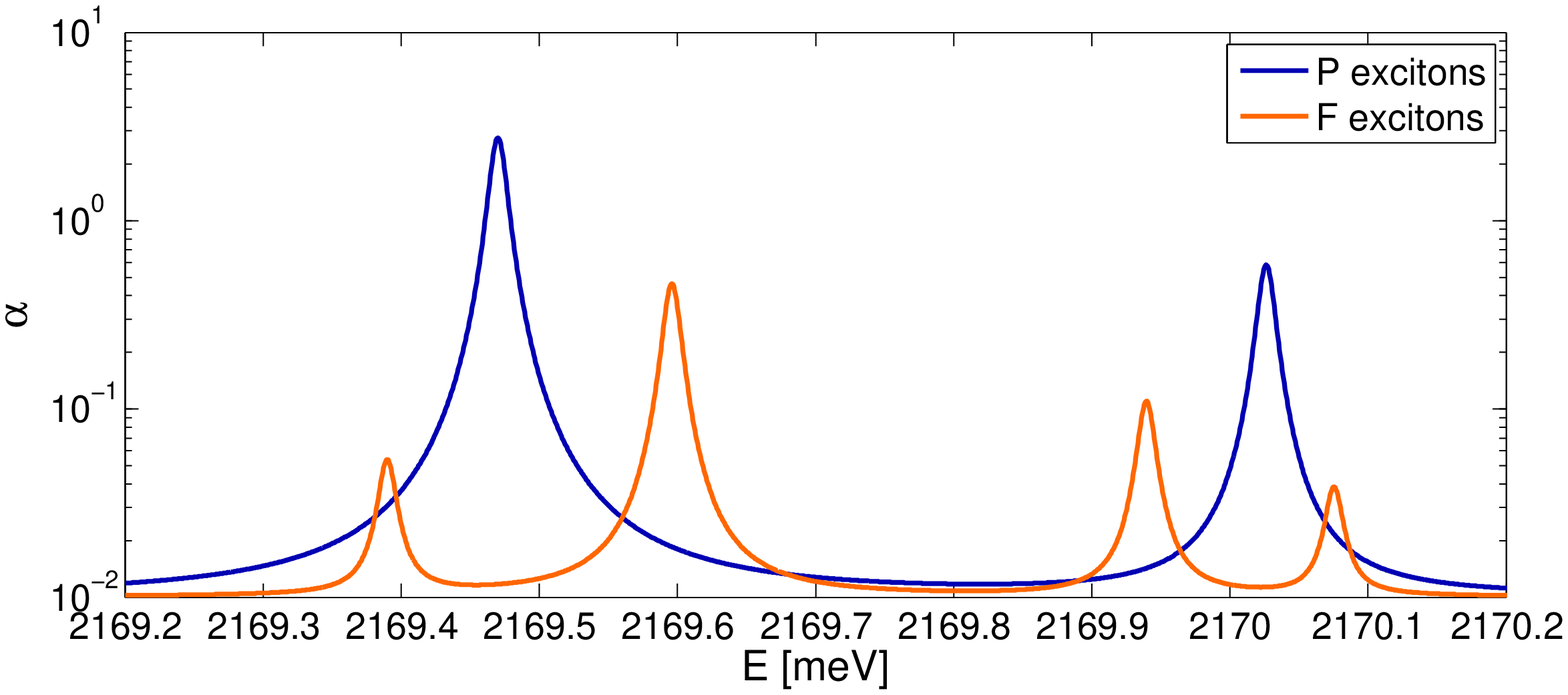}b)
\hfill \caption{a) The same as in Fig.~\ref{fig1_1}, in the
energetic region of $n=7$ exciton, without the electric field, b)
for the field strength 50 V/cm }\label{Fig11}
\end{figure}

\begin{figure}[h]
\includegraphics[width=1\linewidth]{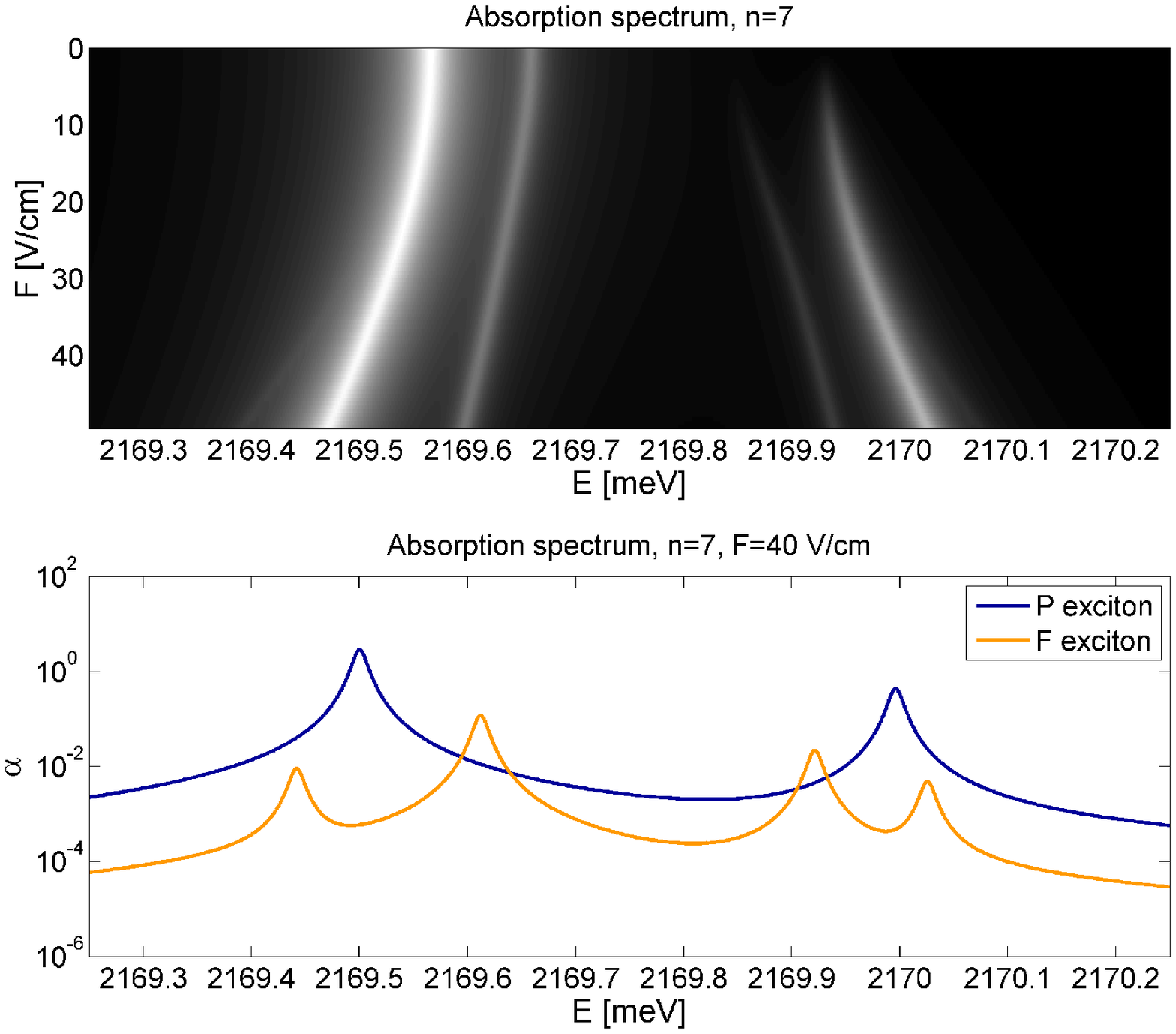}a)
\hfill
\includegraphics[width=0.9\linewidth]{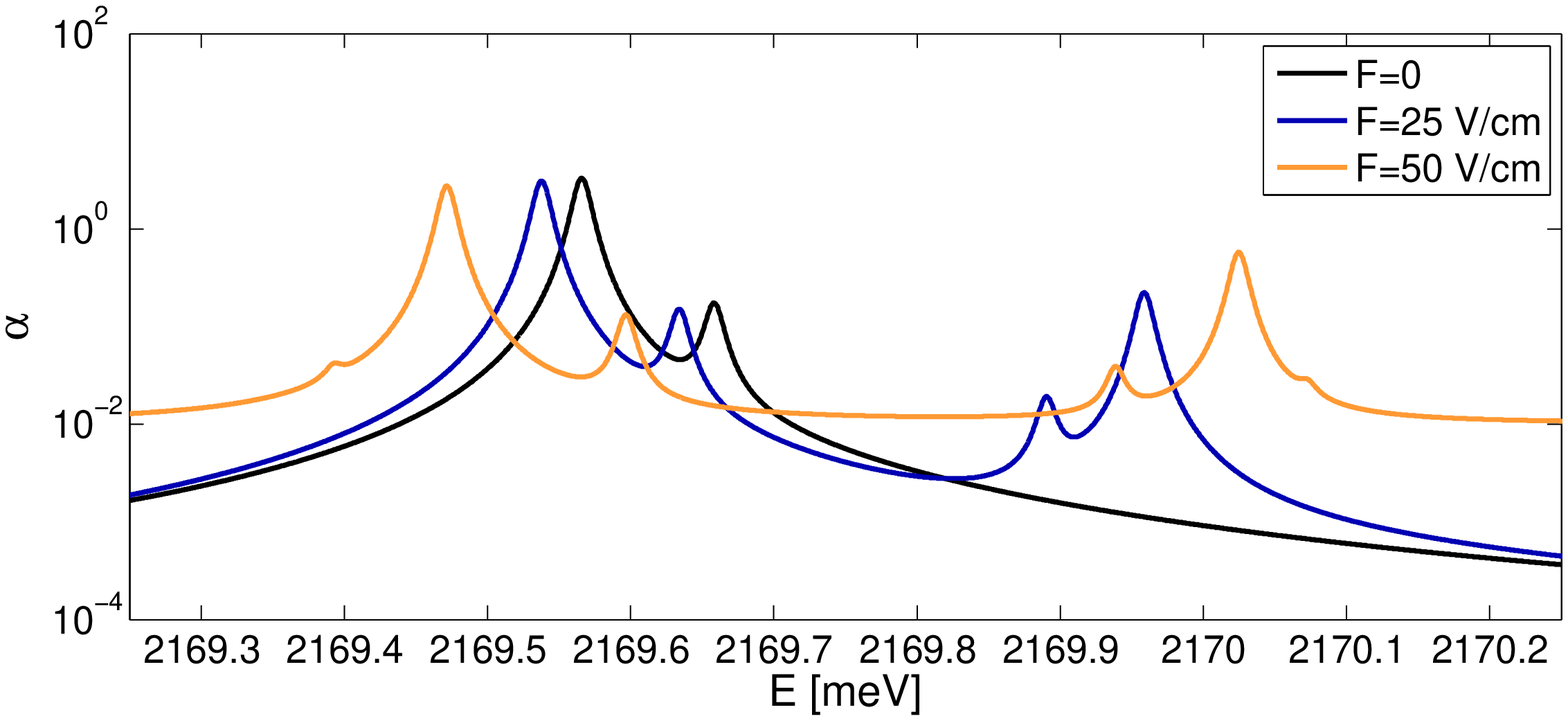}b)
\hfill \caption{a) The changes in the absorption, as a function of
the applied field strength, for the range (0,50) V/cm, b) the same
for three chosen values of the field strength }\label{Fig12}
\end{figure}

\begin{figure}[h]
\includegraphics[width=1\linewidth]{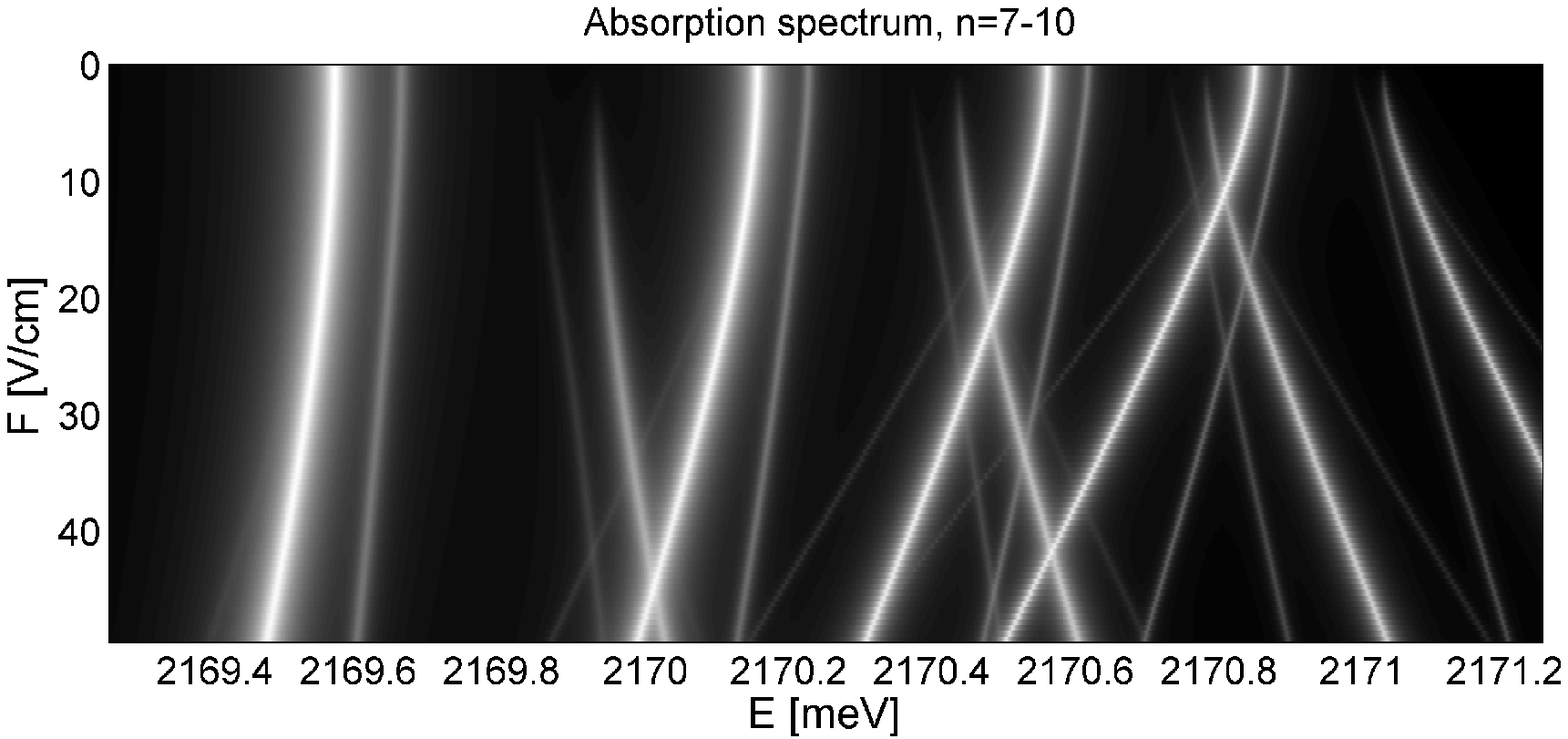}
\caption{Absorption spectrum of Cu$_2$O crystal, in the
energetic region of $n=7-10$ excitonic states as a function of the
applied field strength } \label{Fig13}
\end{figure}

The energies $E_{n\ell m}$ were obtained from the relations
(\ref{energies1}) with the effective Rydberg energy $R^*$ and
mass-anisotropy parameter $\gamma$. We have used the values
$E_g=2172~\hbox{meV}, R^*=86.981~\hbox{meV}$,
$\Delta^{(2)}_{LT}=10~\mu\hbox{eV}$ which is common value in available literature, $\gamma=0.5351$, and
phenomenological value of damping ${\mit\Gamma}=0.1~\hbox{meV}$.  The results for the absorption,
which seem the most important, are reported in
Figs.~\ref{fig1_1}-\ref{Fig13}.

 In Fig.~\ref{fig1_1} we show the absorption spectrum in the region of $n=4-7$ excitons, for two values of the applied field.
 Since the
absorption peaks decrease quite rapidly  the logarithmic scale is
applied. For clarity, we present in Fig.~\ref{Fig3ab}  the
contributions of $P$ and $F$ excitons separately.
The effects of the applied field are more evident when we display
the difference $\Delta\alpha=\alpha(F)-\alpha(F=0)$. Such
difference, for $F=15~\hbox{V/cm}$, is shown in
Fig.~\ref{Figdiff4}. We observe that the numbers of additional
peaks with increasing distances between them in comparison to the situation
without an electric field. In our model this additional
interaction is included in the matrix elements
$V^{(n)}_{\ell_1-1\ell_1m_1}$ (Eq.~(\ref{V12})), which values
increase with the state number (see Table~\ref{matrix.elements}).
Therefore in the following we will focus our attention on higher
number states. It should be stressed that for these  states
oscillator strengths are strong enough to warrant the robust and
stable structure, which is important for possible further
applications. In Fig.~\ref{Figasess} we show the electroabsorption
in the energetic region of $n=8-10$ excitonic states, for two
values of the applied field strength. The $P$ and $F$ states are
clearly distinguished.
As it was reported in ref.~\cite{Verhandl}, the electric field
strength can reach 50 V/cm so we performed numerical simulations  to examine
the influence of field strength on electrooptical properties of
our system.
 It is visible in Fig.~\ref{Fig11} where one can see the basic
effect of the applied field: the Stark  shift of the main peaks
and the appearance of new resonances, especially evident in the
case of \emph{F} excitons.
The changes in the absorption, as a function of the applied field
strength, for the range (0,50) V/cm and near the $n=7$ state, are
presented in Fig.~\ref{Fig12} a. The absorption shape for three
chosen values of the field are given in Fig.~\ref{Fig12} b. The
effect of the Stark shift and changes in the oscillator strength
can be observed.
When we extend the energy interval to include more states, we
observe evident mixing and  overlapping of the lines of the
neighboring  states accompanied by spreading of Stark shifts with
increasing of field strength. (Fig.~\ref{Fig13}). Our theoretical
predictions are very close to the experimental results of Sch\"{o}ne et
al \cite{Schoene}.

\begin{figure}
\includegraphics[width=1\linewidth]{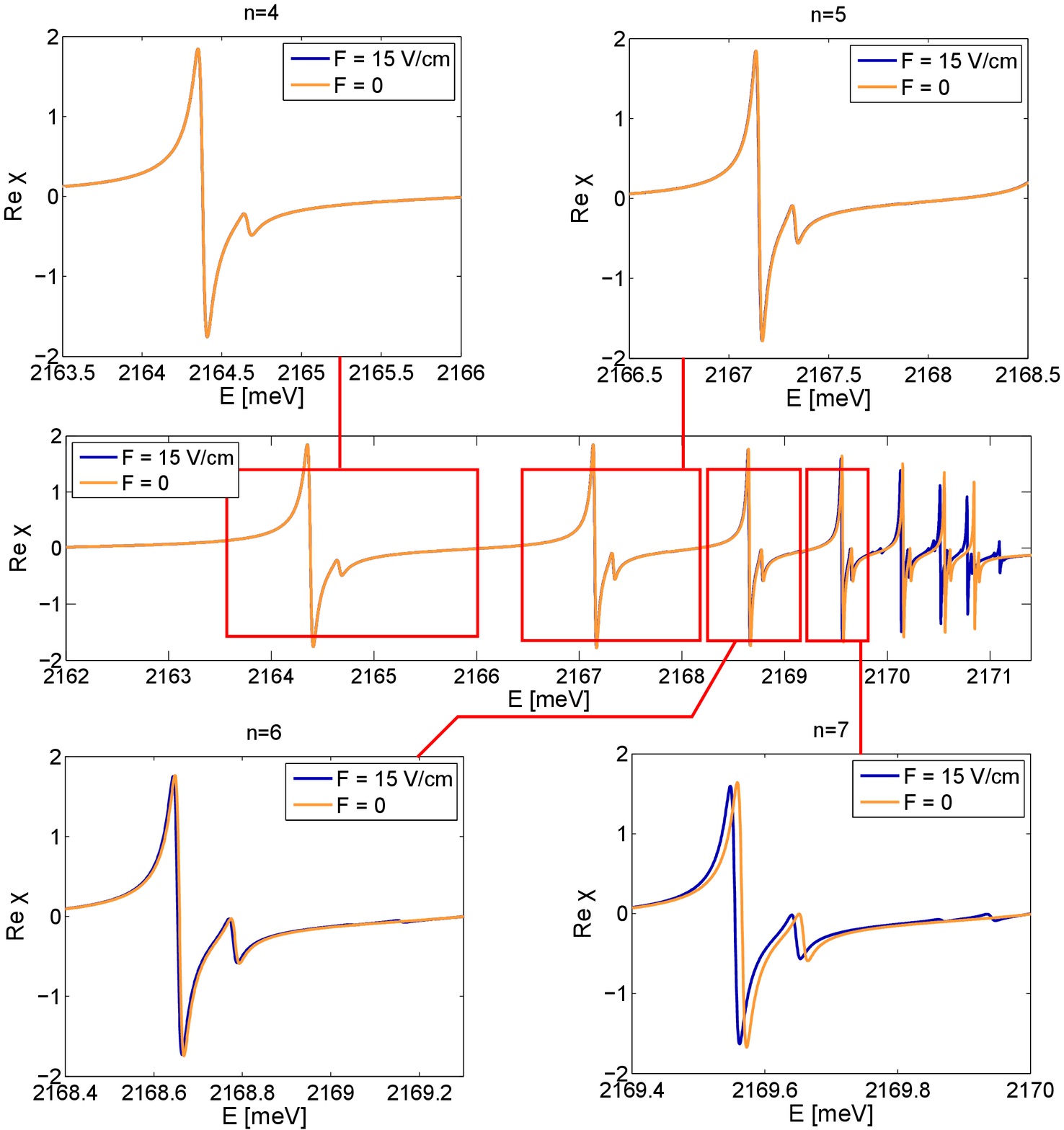}
\caption{The real part of susceptibility of Cu$_2$O crystal in the
energetic region of $n=4-10$ excitonic states, for the electric
field $F=15~\hbox{V/cm}$ and $F=0$. The logarithmic scale is
applied. Insets show absorption spectrum near selected
states.}\label{Fig9}
\end{figure}

\begin{figure}[!htb]
\begin{samepage}
\includegraphics[width=1\linewidth]{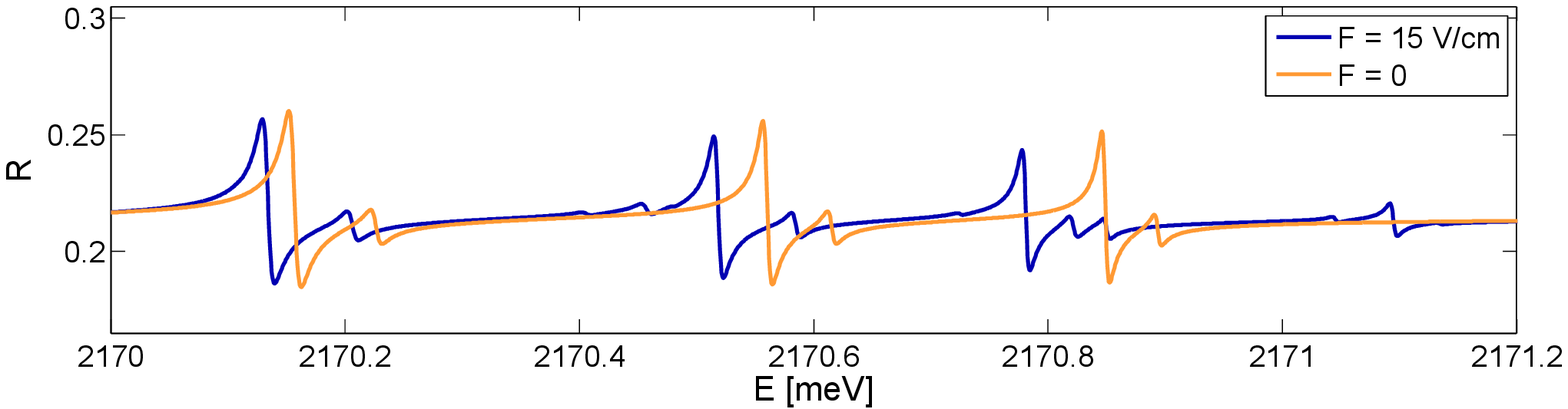}a)
\includegraphics[width=1\linewidth]{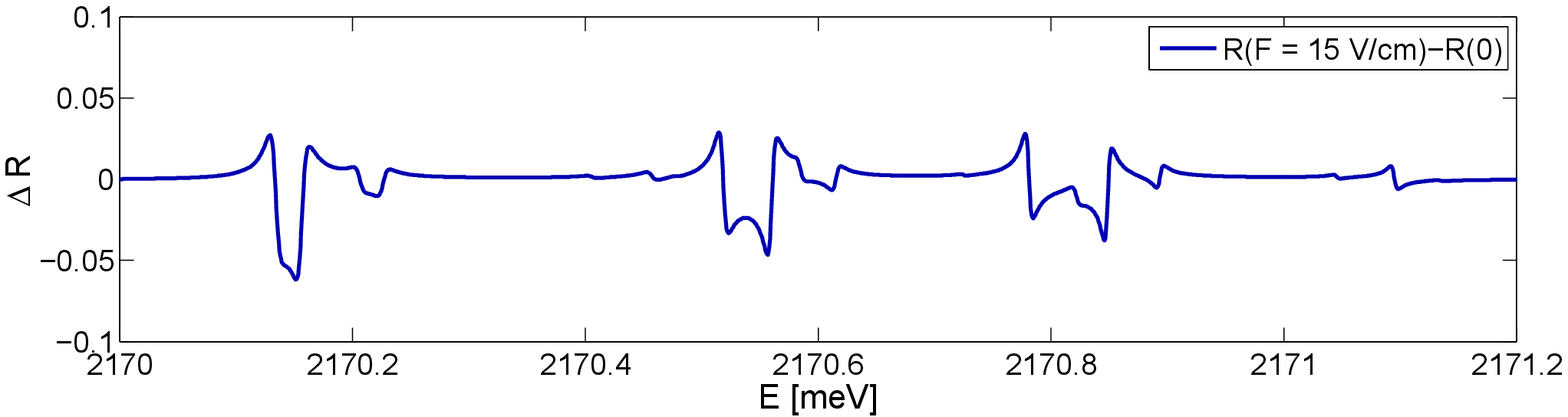}b)
\includegraphics[width=1\linewidth]{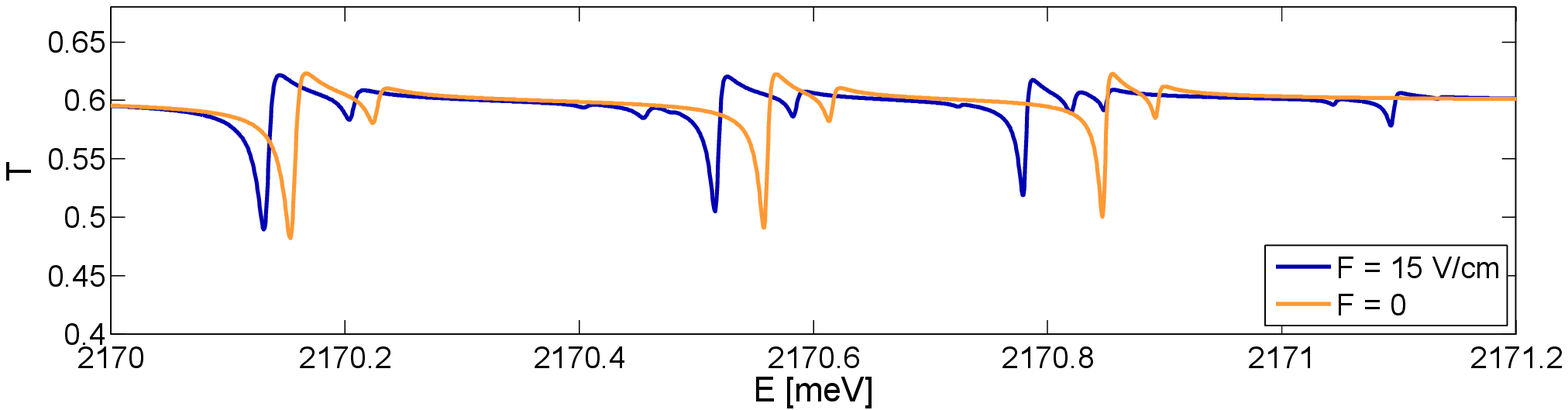}c)
\includegraphics[width=1\linewidth]{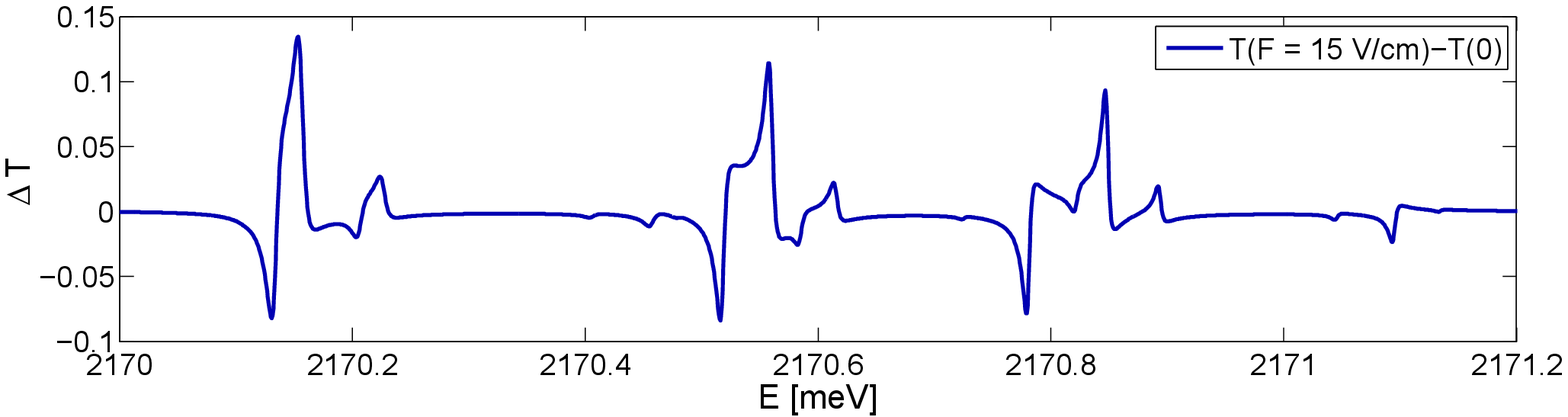}d)
\caption{a) The reflection coefficient of a Cu$_2$O
crystal, in the energetic region of $n=8-10$ excitonic states, for
two values of the electric field. b) The difference $\Delta R=R(F)-R(F=0)$ for $F=15~\hbox{V/cm}$. c) The transmissivity $T$, and d) $\Delta T=T(F)-T(0)$, for $F=15~\hbox{V/cm}$.} \label{Fig5}
\end{samepage}
\end{figure}

 Our method
allows to calculate both the real and the imaginary part of the
susceptibility, without using the Kramers-Kronig relations. The
results for the real part of $\chi$ are presented in Fig. \ref{Fig9}. Having the real and imaginary part of the
susceptibility, we have been able to get the effective dielectric
function from (\ref{epsiloneffective}) and other optical
functions, in particular, the reflection coefficient
(Eq.~(\ref{Rcoeff})), which is shown in Fig.~\ref{Fig5} a). Its shape
resembles the real part of the susceptibility. We notice the red
shift of the main peaks, changes in the oscillator strength,
appearance of new peaks when the field is applied, and decreasing 
the effects for the energies above the 2.171 eV. Similar as it was
done for the electroabsorption (Fig.~\ref{Figdiff4}), we plot the
difference $\Delta R=R(F)-R(0)$ for the energetic region of
$n=8-10$ excitonic resonances (Fig.~\ref{Fig5} b). It can be seen
that the electrooptical effects are noticeable, maxima of reflectivity are back-shifted
 and due to electric field new peaks have occurred.

Finally, making use of Eq. (\ref{transmission}), we have
calculated the transmissivity of the considered above Cu$_2$O
crystal, taking the size $L=30~\mu\hbox{m}$. The results for the
transmissivity $T$ and the difference $\Delta T=T(F)-T(0)$ are
shown in Fig.~\ref{Fig5} c,d. The same tendency as for reflectivity can be observed.

\section{Conlusions}\label{conclusions}
The main results of our paper can be summarized as follows. We
have proposed a procedure based on the RDMA approach that allows
to obtain analytical expressions for the electrooptical functions
of semiconductor crystals including high number Rydberg excitons.
Our results have general character because arbitrary exciton
angular momentum number and arbitrary applied field strength are
included. We have chosen the example of cuprous dioxide, inspired
by the recent experiment by
 Kazimierczuk \emph{et al}~\cite{Kazimierczuk}. We have calculated the electrooptical functions (susceptibility, absorption, reflection, and transmission),
  obtaining a  good agreement between the calculated and the experimentally observed
  spectra. Our results  confirm the fundamental  peculiarity
of Stark effect - shifting, splitting and, as a result for higher excitonic states, mixing of spectral lines.
   In particular, we obtained the splitting of P and F excitons, with increasing number of peaks corresponding to  increasing state number. We could assess the observed
peaks to excitonic states, which are symmetry forbidden when the
electric field is absent. On the basis of our theory we have
predicted the range of energy where one could observe the Stark
splitting and shifting for Rydberg excitons. All these interesting
features of excitons with high $n$ number which are examined
     and discussed  on the basis on our theory might possibly provide deep insight into the nature of Rydberg excitons
     in solids and provoke their application to design all-optical flexible switchers and future implementation in
      quantum information processing. Rydberg excitons in  cuprous oxide are also promising candidates for observing the influence of magnetic fields effects.
Very recently the transmission spectrum of Cu$_2$O yellow series
was registered in magnetic fields for states with high \emph{n}
number showing extraordinary complex splitting pattern of
levels.\cite{assmann_2016} The approach similar to described above
could be used to analyze this experiment.

\appendix
\section{Anisotropic Schr\"{o}dinger equation}\label{Appendix A}
Below we follow the calculations from ref. \cite{Zeitschrift},
correcting and supplementing them. Consider a two band
semiconductor with an isotropic conduction band (electron) mass
$m_e$ and anisotropic hole mass with the components
$m_{h\parallel}, m_z$, with corresponding reduced masses
\begin{equation}
\frac{1}{\mu_\parallel}=\frac{1}{m_e}+\frac{1}{m_{h\parallel}},\qquad
\frac{1}{\mu_z}=\frac{1}{m_e}+\frac{1}{m_{hz}}.
\end{equation}
Anisotropic Schr\"{o}dinger equation for the relative
electron-hole motion, with the above reduced masses and with a
screened Coulomb interaction, has the form
\begin{eqnarray}\label{anisotropic1}
&&\biggl[-\frac{\hbar^2}{2\mu_\parallel}\left(\frac{\partial^2}{\partial
x^2}+\frac{\partial^2}{\partial
y^2}\right)-\frac{\hbar^2}{2\mu_z}\frac{\partial^2}{\partial
z^2} \nonumber \\
&&-\frac{e^2}{4\pi\epsilon_0\epsilon_b\sqrt{x^2+y^2+z^2}}\biggr]\psi=E\psi.
\end{eqnarray}
Using scaled variables
\begin{eqnarray}
&&x=\xi a^*,\qquad y=\eta a^*,\qquad z=\zeta \gamma a^*,\quad
\gamma=\sqrt{\mu_\parallel/\mu_z},\nonumber\\
&&a^{*-1}=\frac{\mu_\parallel}{\hbar^2}\frac{e^2}{4\pi\epsilon_0\epsilon_b},\qquad
\frac{2\mu_\parallel}{\hbar^2}a^{*2}=\frac{1}{R^*},
\end{eqnarray}
we transform Eq. (\ref{anisotropic1}) into
\begin{equation}
\left(\nabla_{\rho^2}-\frac{2}{\rho\sqrt{\sin^2\theta+\gamma^2\cos^2\theta}}\right)\psi(\rho,\theta,\phi)=\varepsilon\psi(\rho,\theta,\phi)
\end{equation}
where $\rho=\sqrt{\xi^2+\eta^2+\zeta^2},\qquad
\varepsilon=E/R^*,$ $\nabla_{\rho^2}$ is the common
Laplace operator in spherical coordinates.
 We are looking for the solution in the form
 \begin{equation}
\psi(\rho,\theta,\phi)=\sum\limits_{\ell} R_\ell (\rho)Y_{\ell
m}(\theta,\phi).
\end{equation}
 Multiplicating both sides with $Y^*_{\ell'm'}$,
integrating and taking into account only the diagonal terms
$\ell=\ell', m=m'$ we obtain the following equation for the radial
part $R$
\begin{equation}\label{rho}
\left[\frac{{\rm d}^2}{{\rm d}\rho^2}+\frac{2}{\rho}\frac{\rm
d}{{\rm d}\rho}-\kappa^2+\frac{2}{\rho}\eta_{\ell
m}-\frac{\ell(\ell+1)}{\rho^2}\right] R=0,
\end{equation}
with
\begin{equation}\label{etaellm}
\eta_{\ell m}=\int\limits_0^{2\pi} {\rm d}\phi\int\limits_0^\pi
\sin\theta\,{\rm d}\theta\frac{\left|Y_{\ell
m}\right|^2}{\sqrt{\sin^2\theta +\gamma^2 \cos^2\theta}},
\end{equation}
and with $\kappa^2=-\varepsilon$, assuming that we consider only
the bound states. Some values for $\eta_{\ell m}$ were given in
Refs. \cite{Zielinska.PRB} and \cite{Zeitschrift}. Mostly $\gamma$
is close to 1. In this case, to a good approximation
\begin{equation}
\eta_{\ell
m}=1+\frac{(1-\gamma)(2\ell^2+2\ell-1-2m^2)}{2(2\ell-1)(2\ell+3)}.
\end{equation}
Making use of substitutions
\begin{equation}\label{relations}
z=2\kappa\rho,\quad \lambda=\frac{\eta_{\ell m}}{\kappa},\quad
F=R(z)z,\quad \mu=\ell+\frac{1}{2},
\end{equation}
we transform (\ref{rho}) into the equation
\begin{equation}\label{Whittaker}
\frac{{\rm d}^2}{{\rm
d}z^2}F+\left(-\frac{1}{4}+\frac{\lambda}{z}+\frac{(1/4)-\mu^2}{z^2}\right)F=0.
\end{equation}
 The above equation has two linearly independent solutions
$M_{\lambda,\mu}, W_{\lambda,\mu}$ known as the Whittaker
functions. They are related to the more familiar Kummer functions
(confluent hypergeometric functions) by the relations
\begin{eqnarray*}
M_{\lambda,\mu}(z)=z^{\mu+1/2}e^{-z/2}M\left(\mu-\lambda+\frac{1}{2},2\mu+1,z\right),\nonumber\\
W_{\lambda,\mu}(z)=z^{\mu+1/2}e^{-z/2}U\left(\mu-\lambda+\frac{1}{2},2\mu+1,z\right).
\end{eqnarray*}
We choose the function $M$ which is finite for $z=0$ and, with
respect to the relations (\ref{relations}) we obtain the radial
part $R$ in the form
\begin{equation}\label{radial1}
R=N(2\kappa\rho)^{\ell}e^{-\kappa\rho}M\left(\ell+1-\lambda,2\ell+2,2\kappa\rho\right),
\end{equation}
$N$ being the normalization constant. The function $R$ is finite
for $\rho\to\infty$ when the first argument of the Kummer function
is 0 or negative integer. Thus
\begin{equation}
\ell+1-\frac{\eta_{\ell m}}{\kappa}=-N,
\end{equation}
which gives
\begin{equation}
\frac{\eta_{\ell m}}{\kappa}=N+\ell+1=n,
\end{equation}
and, finally \begin{equation}\label{energies}
\varepsilon=-\frac{\eta_{\ell m}^2}{n^2}.
\end{equation}
Inserting the result for $\kappa$ into (\ref{radial1}) we obtain
the radial function $R$ in the form
\begin{eqnarray}\label{radial2}
&&R(\rho)=R_{n\ell m}(\rho)=N_{n\ell m}\left(\frac{2\eta_{\ell
m}\rho}{n}\right)^\ell e^{-\eta_{\ell
m}\rho/n}\nonumber\\
&&\times\,M\left(-n+\ell+1,2\ell+2,\frac{2\eta_{\ell
m}\rho}{n}\right).
\end{eqnarray}
Using the integral (for example, \cite{Landau})
\begin{eqnarray*}
&&J_\nu=\int\limits_0^\infty
e^{-kz}z^{\nu-1}\left[M(-n,\gamma,kz)\right]^2{\rm
d}z\nonumber\\
&&=\frac{\Gamma(\nu)\;n!}{k^\nu\gamma(\gamma+1)\ldots(\gamma+n-1)}\biggl\{1+\frac{n(\gamma-\nu-1)(\gamma-\nu)}{1^2\cdot
\gamma}\nonumber\\
&&+\frac{n(n-1)(\gamma-\nu-2)(\gamma-\nu-1)(\gamma-\nu)(\gamma-\nu+1)}{1^2\cdot
2^2\cdot\gamma(\gamma+1)}+\ldots\biggr\}
\end{eqnarray*}
we obtain the normalization constant in the form
\begin{equation}
N_{n\ell m}=\left(\frac{2\eta_{\ell
m}}{n}\right)^{3/2}\frac{1}{(2\ell
+1)!}\sqrt{\frac{(n+\ell)!}{2n(n-\ell-1)!}}.
\end{equation}
Thus the radial part of the solution of the anisotropic
Schr\"{o}dinger equation has the form
\begin{eqnarray}\label{radialfinal}
&&R_{n\ell m}(\rho)=\left(\frac{2\eta_{\ell
m}}{n}\right)^{3/2}\frac{1}{(2\ell
+1)!}\sqrt{\frac{(n+\ell)!}{2n(n-\ell-1)!}}\nonumber\\
&&\\ &&\times\left(\frac{2\eta_{\ell m}\rho}{n}\right)^\ell
e^{-\eta_{\ell m}\rho/n}M\left(-n+\ell+1,2\ell+2,\frac{2\eta_{\ell
m}\rho}{n}\right).\nonumber
\end{eqnarray}
Using the relation
$$L_N^\alpha(x)={N+\alpha\choose N}M(-N,\alpha+1,x)$$
between the Kummer function and the Laguerre polynomials, we can
express the radial function $R$ in terms of them
\begin{eqnarray}
&&R_{n\ell m}(\rho)=\left(\frac{2\eta_{\ell
m}}{n}\right)^{3/2}\sqrt{\frac{(n-\ell-1)!}{2n(n+\ell)!}}\left(\frac{2\eta_{\ell
m}\rho}{n}\right)^\ell\nonumber\\
&&\times L_{n-\ell-1}^{2\ell+1}\left(\frac{2\eta_{\ell
m}\rho}{n}\right) e^{-\eta_{\ell m}\rho/n}.
\end{eqnarray}
\section{Derivation of the expansion coefficients}\label{Appendix B}
Inserting the expansion (\ref{expansion}) into
(\ref{constitutiveeqn}) and making use of the relations
\begin{eqnarray}
\cos\theta\,Y_{\ell m}&=&\sqrt{\frac{(\ell
+1+m)(\ell+1-m)}{(2\ell+1)(2\ell+3)}}Y_{\ell+1
m}\nonumber\\
&+&\sqrt{\frac{(\ell +m)(\ell-m)}{(2\ell+1)(2\ell-1)}}Y_{\ell-1
m},\nonumber\\
\langle \ell_1m_1\vert\cos\theta\vert \ell_2 m_2\rangle&\neq&
0\quad\hbox{ if}\quad m_1=m_2~\hbox{and}~\ell_1=\ell_2\pm
1\quad\nonumber\\
\langle \ell m\vert\cos\theta\vert\ell -1
m\rangle&=&\sqrt{\frac{\ell^2-m^2}{4\ell^2-1}},
\end{eqnarray}
and making use of the orthogonality properties of the
eigenfunctions $R_{n\ell}, Y_{\ell m}$ we obtain the system of
equations (\ref{basic}) for the expansion coefficients. The
equations (\ref{basic}) form, in general, an infinite system of
linear equations. Therefore a certain cut-off must be applied.
Having in mind the properties of Cu$_2$O we put $n=n_1$, i.e. we
neglect the interaction between the states with different quantum
number $n$. It is due to the fact that the energy differences
between the states are much larger than the perturbations caused
by the electric field. In consequence, the infinite system of
equations is reduced to a set of subsystems of equations for each
value of $n$. The subsystems consist, in general, of $2n^2$
equations labeled by different values of $\ell$ and $m$. With
respect to the properties of Cu$_2$O, we will consider the $P$
excitons ($\ell=1$) and $F$ excitons ($\ell=3$). The lowest $P$
exciton state is given by $n=2, \ell=1, m=0$. From (\ref{basic}),
with $\textbf{M}$ given~by~\cite{Zielinska.PRB}

\begin{eqnarray}\label{gestoscwzbronione}
{\bf M}({\bf r})&=&
\textbf{e}_r\,M_{10}\frac{r+r_0}{2r^2r_0^2}e^{-r/r_0}=\textbf{e}_r
M(r) =\nonumber\\&=&\textbf{i}M_{10}\frac{r+r_0}{4{\rm
i}r^2r_0^2}\sqrt{\frac{8\pi}{3}}\left(Y_{1,-1}-Y_{1,1}\right)e^{-r/r_0}+\nonumber\\
&+&\textbf{j}M_{10}\frac{r+r_0}{4r^2r_0^2}\sqrt{\frac{8\pi}{3}}\left(Y_{1,-1}+Y_{1,1}\right)e^{-r/r_0}+\nonumber\\&+&\textbf{k}M_{10}\frac{r+r_0}{2r^2r_0^2}
\sqrt{\frac{4\pi}{3}}Y_{10}e^{-r/r_0}, \end{eqnarray}and its
$Y_{10}$ component, one obtains 4 equations
\begin{eqnarray}
\begin{array}{rcccccc}\nonumber
W_{200}c_{200}&+&V^{(2)}_{010}c_{210}&&&=&0\\
V^{(2)}_{010}c_{200}&+&W_{210}c_{210}&&&=&X_{210}\\
&&&W_{211}c_{211}&&=&0\\
&&&&W_{21-1}c_{21-1}&=&0
\end{array}\\
\end{eqnarray}
where we took only the allowed combinations for the $n,\ell,m$.
Thus we obtain
\begin{equation}\label{c21}
c_{210}=C_{210}X_{210}=\frac{W_{200}X_{210}}{W_{200}W_{210}-\left(V^{(2)}_{010}\right)^2}.
\end{equation}
For the $P,n=3$ exciton state we use $n^2=9$ combinations:
\begin{eqnarray}\label{enumeration}
\begin{array}{rcccccc}
300(x_1)&310(x_2)&311(x_3)&31-1(x_4)&&&\\
320(x_5)&321(x_6)&32-1(x_7)&&&&\\
322(x_8)&32-2(x_9)&&&&&\\
\end{array}
\end{eqnarray}
and obtain equations
\begin{widetext}
\begin{eqnarray}\label{eqn3P}
\begin{array}{rccccccccc}
W_{300}x_1&+&V_{010}^{(3)}x_2&+\ldots&&&&&=&0\\
V_{010}^{(3)}x_1&+&W_{310}x_2&+\ldots&&+V_{120}^{(3)}x_5&&&=&X_{310}\\
&&&W_{311}x_3&&\ldots&+V_{121}^{(3)}x_6&&=&0\\
&&&&W_{31-1}x_4&&\ldots&+V_{12-1}^{(3)}x_7&=&0\\
&&V_{120}^{(3)}x_2&&\ldots&+W_{320}x_5&&&=&0\\
&&&V_{121}^{(3)}x_3&&\ldots&+W_{321}x_6&&=&0\\
&&&&V_{12-1}^{(3)}x_4&&\ldots&+W_{32-1}x_7&=&0\\
\end{array}
\end{eqnarray}
where $c_{300}=x_1, c_{310}=x_2$ etc., with regard to
(\ref{enumeration}). For remaining the remaining coefficients we
have $x_8=x_9=0$. The resulting coefficient
$c_{310}=C_{310}X_{310}$  is given by the formula (\ref{c310}). In
the expressions for $C_{210}$ and $C_{n10}$ one can separate the
real and imaginary part, obtaining
\begin{eqnarray}\label{realimaginar}
C_{n10}&=&
\frac{(E_{Tn00}-E)\left[\left(E_{Tn00}-E\right)\left(E_{Tn10}-E\right)-\left(V_{010}^{(n)}\right)^2-{\mit\Gamma}^2\right]+\left(E_{Tn10}
+E_{Tn00}-2E\right){\mit\Gamma}^2}{\left[\left(E_{Tn00}-E\right)\left(E_{Tn10}-E\right)-\left(V_{010}^{(n)}\right)^2-{\mit\Gamma}^2\right]^2+\left(E_{Tn10}
+E_{Tn00}-2E\right)^2{\mit\Gamma}^2}\nonumber\\ &+\;{\rm
i}{\mit\Gamma}&\frac{\left(E_{Tn00}-E\right)^2+\left(V_{010}^{(n)}\right)^2+{\mit\Gamma}^2}{\left[\left(E_{Tn00}-
E\right)\left(E_{Tn10}-E\right)-\left(V_{010}^{(n)}\right)^2-{\mit\Gamma}^2\right]^2+\left(E_{Tn10}
+E_{Tn00}-2E\right)^2{\mit\Gamma}^2}\;.
\end{eqnarray}
Clearly $\hbox{Im}\;C_{n10}>0$. Introducing notation
\begin{eqnarray}\label{definitions}
E_\ell^{(n)}&=&E_{Tn\ell 0}-E,\nonumber\\
W_{n\ell 0}(k=0)&=&E_\ell^{(n)}-{\rm i}{\mit\Gamma},\\
R_{\ell\ell_1}^{(n)}&=&E_\ell^{(n)}E_{\ell_1}^{(n)}-{\mit\Gamma}^2,\nonumber\\
S_{\ell\ell_1}^{(n)}&=&E_\ell^{(n)}+E_{\ell_1}^{(n)},\nonumber
\end{eqnarray}
we put eq. (\ref{realimaginar}) into a more compact form
\begin{eqnarray}\label{realimaginarcompact}
C_{n10}&=&
\frac{E_0^{(n)}\left[R_{01}^{(n)}-\left(V_{010}^{(n)}\right)^2\right]+S_{01}^{(n)}{\mit\Gamma}^2}
{\left[R_{01}^{(n)}-\left(V_{010}^{(n)}\right)^2\right]^2+\left(S_{01}^{(n)}\right)^2{\mit\Gamma}^2}
+\;{\rm
i}{\mit\Gamma}\frac{\left(E_0^{(n)}\right)^2+\left(V_{010}^{(n)}\right)^2+{\mit\Gamma}^2}
{\left[R_{01}^{(n)}-\left(V_{010}^{(n)}\right)^2\right]^2+\left(S_{01}^{(n)}\right)^2{\mit\Gamma}^2}\;.
\end{eqnarray}
For $F$ excitons, when $n_1\geq 4, \ell=3,2,1,0, m=0$, we obtain
the following equations for the expansion coefficients
\begin{eqnarray}
\begin{array}{cccccc}
c_{n_130}W_{n_130}&+c_{n_120}V^{(n_1)}_{230}&&&=X_{n_130},\\
c_{n_130}V^{(n_1)}_{230}&+c_{n_120}W_{n_120}&+c_{n_110}V_{120}^{(n_1)}&&=0,\\
&c_{n_120}V^{(n_1)}_{120}&+c_{n_110}W_{n_110}&+c_{n_100}V_{010}^{(n_1)}&=0,\\
&&c_{n_110}V_{010}^{(n_1)}&+c_{n_100}W_{n_100}&=0.
\end{array}
\end{eqnarray}
with the result for the relevant coefficient
\begin{equation}\label{cn130}
c_{n_130}=X_{n_130}C_{n_130}=X_{n_130}\frac{W_{n_120}W_{n_110}W_{n_100}-W_{n_120}\left(V_{010}^{(n_1)}\right)^2
-W_{n_100}\left(V_{120}^{(n_1)}\right)^2}{\Delta},
\end{equation}
where
\begin{eqnarray*}
&&\Delta=\left|\begin{array}{cccccc}
W_{n_130}&V^{(n_1)}_{230}&0&0&\\
V^{(n_1)}_{230}&W_{n_120}&V_{120}^{(n_1)}&0&\\
0&V^{(n_1)}_{120}&W_{n_110}&V_{010}^{(n_1)}&\\
0&0&V_{010}^{(n_1)}&W_{n_100}&\\
\end{array}\right|\nonumber\\
&&=W_{n_130}\left[W_{n_120}W_{n_110}W_{n_100}-W_{n_120}\left(V_{010}^{(n_1)}\right)^2
-W_{n_100}\left(V_{120}^{(n_1)}\right)^2\right]\\
&&-\left(V^{(n_1)}_{230}\right)^2\left[W_{n_110}W_{n_100}-\left(V_{010}^{(n_1)}\right)^2\right].
\end{eqnarray*}
Using the definitions (\ref{definitions}), $C_{n_130}$ can be put
into the form \begin{eqnarray}\label{cn1301}
C_{n_130}&=&\frac{ac-bd}{a^2+b^2}+{\rm
i}\;\frac{ad+bc}{a^2+b^2},\nonumber\\
a&=&R^{(n_1)}_{01}R^{(n_1)}_{23}+\left(V_{010}^{(n_1)}\right)^2\left(V^{(n_1)}_{230}\right)^2-{\mit\Gamma}^2S_{01}^{(n_1)}S_{23}^{(n_1)}
-R^{(n_1)}_{01}\left(V^{(n_1)}_{230}\right)^2\nonumber\\
&&-R^{(n_1)}_{03}\left(V^{(n_1)}_{120}\right)^2-R^{(n_1)}_{23}\left(V^{(n_1)}_{010}\right)^2,\\
b&=&{\mit\Gamma}\left[R^{(n_1)}_{01}S_{23}^{(n_1)}+R^{(n_1)}_{23}S_{01}^{(n_1)}-S_{01}^{(n_1)}\left(V^{(n_1)}_{230}\right)^2-
S_{03}^{(n_1)}\left(V_{120}^{(n_1)}\right)^2-S_{23}^{(n_1)}\left(V_{010}^{(n_1)}\right)^2\right]\nonumber\\
c&=&E^{(n_1)}_2R^{(n_1)}_{01}-{\mit\Gamma}^2S_{01}^{(n_1)}-E^{(n_1)}_2
\left(V_{010}^{(n_1)}\right)^2-E^{(n_1)}_0\left(V_{120}^{(n_1)}\right)^2,\nonumber\\
d&=&{\mit\Gamma}\left[\left(V_{010}^{(n_1)}\right)^2+\left(V_{120}^{(n_1)}\right)^2-S_{01}^{(n_1)}E^{(n_1)}_2-R^{(n_1)}_{01}\right].\nonumber
\end{eqnarray}
For $F$ excitons, when $n\geq 5$, we can extend the basis taking
$\ell=4,3,2,1,0$ and $m=0$, obtaining
\begin{eqnarray}\label{cn130ext}
c_{n_130}&=&X_{n_130}\frac{W_{n_140}\left[W_{n_120}W_{n_110}W_{n_100}-W_{n_120}\left(V_{010}^{(n_1)}\right)^2-
W_{n_100}\left(V_{120}^{(n_1)}\right)^2\right]}{\Delta},\nonumber\\
\Delta&=&\left[W_{n_140}W_{n_130}-\left(V_{340}^{(n_1)}\right)^2\right]\left[W_{n_120}W_{n_110}W_{n_100}-W_{n_120}\left(V_{010}^{(n_1)}\right)^2-
W_{n_100}\left(V_{120}^{(n_1)}\right)^2\right]\nonumber\\
&&-W_{n_140}\left(V^{(n_1)}_{230}\right)^2\left[W_{n_110}W_{n_100}-\left(V_{010}^{(n_1)}\right)^2\right].
\end{eqnarray}
\section{Derivation of the matrix elements $V^{(n)}_{\ell_1-1\ell_1m_1}$}\label{Appendix C}
The matrix elements follow  from the definitions (\ref{V12}):
\begin{eqnarray}\label{V12a1}
&&V^{(n_1)}_{\ell_1-1\ell_1m_1}=eFa^*\sqrt{\frac{\ell_1^2-m_1^2}{4\ell_1^2-1}}\left(\frac{2}{n_1}\right)^3\int\,\rho^3{\rm
d}\rho\,e^{-2\rho/n_1}\biggl\{
\sqrt{\frac{(n_1-\ell_1-1)!}{2n_1(n_1+\ell_1)!}}\left(\frac{2\rho}{n_1}\right)^{\ell_1}L^{2\ell_1+1}_{n_1-\ell_1-1}\nonumber\\
&&\times
\sqrt{\frac{(n_1-\ell_1)!}{2n_1(n_1+\ell_1-1)!}}\left(\frac{2\rho}{n_1}\right)^{\ell_1-1}L^{2\ell_1-1}_{n_1-\ell_1}\biggr\}
\end{eqnarray}
\begin{eqnarray}\label{V12a2}
&&V^{(n_1)}_{\ell_1\ell_1+1m_1}=eFa^*\sqrt{\frac{(\ell_1
+m_1+1)(\ell_1-m_1+1)}{(2\ell_1+1)(2\ell_1+3)}}\left(\frac{2}{n_1}\right)^3\\
&&\times\int\,\rho^3{\rm
d}\rho\,e^{-2\rho/n_1}\biggl\{\sqrt{\frac{(n_1-\ell_1-2)!}{2n_1(n_1+\ell_1+1)!}}\left(\frac{2\rho}{n_1}\right)^{\ell_1}L^{2\ell_1+1}_{n_1-\ell_1-1}
\sqrt{\frac{(n_1-\ell_1-1)!}{2n_1(n_1+\ell_1-1)!}}\left(\frac{2\rho}{n_1}\right)^{\ell_1+1}L^{2\ell_1+3}_{n_1-\ell_1-2}\biggr\}.\nonumber
\end{eqnarray}
with the Laguerre polynomials $L^\alpha_n(x)$ (see
(\ref{laguerre}). Substituting $x=\frac{2\rho}{n_1}$ and  treating
$eFa^*$ as unit, we obtain
\begin{eqnarray}\label{V1ax}
&&V^{(n_1)}_{\ell_1-1\ell_1m_1}=\sqrt{\frac{\ell_1^2-m_1^2}{4\ell_1^2-1}}\left(\frac{n_1}{2}\right)\int\,x^3{\rm
d}x\,e^{-x}x^{2\ell_1-1}\biggl\{
\sqrt{\frac{(n_1-\ell_1-1)!}{2n_1(n_1+\ell_1)!}}L^{2\ell_1+1}_{n_1-\ell_1-1}(x)\nonumber\\
&&\phantom{aaaaaaaaaaaa}\times
\sqrt{\frac{(n_1-\ell_1)!}{2n_1(n_1+\ell_1-1)!}}L^{2\ell_1-1}_{n_1-\ell_1}(x)\biggr\}\nonumber\\
&&=\sqrt{\frac{(\ell_1^2-m_1^2)}{16(4\ell_1^2-1)}\frac{(n_1-\ell_1-1)!}{(n_1+\ell_1)!}\frac{(n_1-\ell_1)!}{(n_1+\ell_1-1)!}}\nonumber\\
&&\phantom{aaaaaaaaaaaa}\times \int\,{\rm
d}x\,e^{-x}x^{2\ell_1+2}L^{2\ell_1+1}_{n_1-\ell_1-1}(x)L^{2\ell_1-1}_{n_1-\ell_1}(x),
\end{eqnarray}
\begin{eqnarray}\label{V12a2x}
&&V^{(n_1)}_{\ell_1\ell_1+1m_1}=\sqrt{\frac{(\ell_1
+m_1+1)(\ell_1-m_1+1)}{(2\ell_1+1)(2\ell_1+3)}}\left(\frac{2}{n_1}\right)^3\nonumber\\
&&\times\int\,\rho^3{\rm
d}\rho\,e^{-2\rho/n_1}\biggl\{\sqrt{\frac{(n_1-\ell_1-2)!}{2n_1(n_1+\ell_1+1)!}}
\left(\frac{2\rho}{n_1}\right)^{\ell_1}L^{2\ell_1+1}_{n_1-\ell_1-1}
\sqrt{\frac{(n_1-\ell_1-1)!}{2n_1(n_1+\ell_1)!}}\left(\frac{2\rho}{n_1}\right)^{\ell_1+1}L^{2\ell_1+3}_{n_1-\ell_1-2}\biggr\}\nonumber\\
&&=\sqrt{\frac{(\ell_1
+m_1+1)(\ell_1-m_1+1)}{16(2\ell_1+1)(2\ell_1+3)}\frac{(n_1-\ell_1-2)!}{(n_1+\ell_1+1)!}\frac{(n_1-\ell_1-1)!}{(n_1+\ell_1)!}}\nonumber\\
&&\phantom{aaaaaaaaaaaa}\times \int\,{\rm
d}x\,e^{-x}x^{2\ell_1+4}L^{2\ell_1+1}_{n_1-\ell_1-1}(x)L^{2\ell_1+3}_{n_1-\ell_1-2}(x).
\end{eqnarray}
In particular, for $V^{(n)}_{010}$ and  in units $eFa^*$, one
obtains
\begin{eqnarray}\label{v010}
V^{(n)}_{010}&=&\frac{1}{\sqrt{3}}\sqrt{\frac{(n-1)!(n-2)!}{16
n!(n+1)!}}\int\limits_0^\infty {\rm d}x\,e^{-x}x^4
L^1_{n-1}(x)L^3_{n-2}(x)\nonumber\\
&=&-\sqrt{\frac{12}{n^2(n^2-1)}}{n\choose n-2}{n+1\choose n-1},
\end{eqnarray}
where we used the following integral involving Laguerre
polynomials \cite{Grad}
\begin{eqnarray}
\int\limits_0^\infty
&&e^{-x}x^{\alpha+\beta}L_m^\alpha(x)L_n^\beta(x){\rm
d}x\nonumber\\
&&=(-1)^{m+n}(\alpha+\beta)!{\alpha+m\choose n}{\beta+n\choose
m}\qquad [\hbox{Re}\,(\alpha+\beta)>-1].
\end{eqnarray}
For $V^{(n)}_{230}$ we have by definition
\begin{eqnarray}\label{v230}
V^{(n)}_{230}&=&\sqrt{\frac{9(n-3)!(n-4)!}{16\cdot
35\,(n+2)!(n+3)!}}\int\limits_0^\infty {\rm
 d}x\,e^{-x}x^8
L^7_{n-4}(x)L^5_{n-3}(x).
\end{eqnarray}
Some numerical values for the elements $V^{(n)}_{010}$ and
$V^{(n)}_{230}$ are given in Table~\ref{matrix.elements}.

Another example, important in view of the formulas (\ref{cc210})
and (\ref{cn130}), will be obtained from eq. (\ref{V12a1}) by
taking $n_1=3, \ell_1=2, m_1=0$

\begin{eqnarray}
V^{(3)}_{120}&=&\sqrt{\frac{4\cdot 0!\cdot 1!}{16\cdot 15\cdot
5!\cdot 4!}}\int\limits_0^\infty {\rm d}x\;
e^{-x}x^6L^5_0(x)L^3_1(x)\nonumber\\
&=&\sqrt{\frac{1}{4\cdot 15\cdot 5!\cdot 4!}}\int\limits_0^\infty
{\rm d}x\; e^{-x}x^6(4-x)=-3\cdot 6!\sqrt{\frac{1}{4\cdot 15\cdot
5!\cdot 4!}}=-3\sqrt{3}\approx -5.196,
\end{eqnarray} since
$L^5_0(x)=1,\;\;L^3_1(x)=4-x.$
\end{widetext}

\begin{table}[ht]
\caption{The matrix elements $V^{(n)}_{010}$, and
$V^{(n)}_{230}\;\;\;$  }\label{matrix.elements}
\begin{tabular}{|c|c|c|c|c|c|} \hline
$n$& 2& 3& 4& 5&6\\\hline
$-V^{(n)}_{010}$ &3.0000 &7.3485 &13.4164 &21.2132 &30.7409\\\hline
$-V^{(n)}_{230}$&&&8.0498&15.2128&23.7144
\\\hline
$n$& 7& 8& 9& 10& \\\hline
$-V^{(n)}_{010}$ &42.0000&54.9909&69.7137&86.1684&\\\hline
$-V^{(n)}_{230}$&33.6749&45.1284&58.0881&72.5603&
\\\hline
\end{tabular}
\end{table}

\end{document}